\documentclass[%
 reprint,
 amsmath,amssymb,
 aps,
floatfix,
]{revtex4-2}
\DeclareUnicodeCharacter{2009}{\,} 
\usepackage{graphicx}
\usepackage{dcolumn}
\usepackage{bm}

\usepackage{graphicx,xcolor}

\date{\today}
\begin{document}
\author{Rupesh Mahore$^{(1)}$}
\author{Oleksandr Gamayun$^{(2)}$}
\author{Guillaume Noetinger$^{(3)}$}
\author{Romain Fleury$^{(3)}$}
\author{Corentin Coulais$^{(2)}$}
\email{coulais@uva.nl}
\author{Benjamin Apffel$^{(4)}$}
\email{benjamin.apffel@ens-lyon.fr}
\affiliation{(1) Institute of Physics, Universiteit van Amsterdam, Science Park 904, 1098 XH Amsterdam, The Netherlands}
\affiliation{ (2) London Institute for Mathematical Sciences, Royal Institution, 21 Albemarle St., London W1S 4BS}
\affiliation{(3) Laboratory of Wave Engineering, EPFL, 1015 Lausanne, Switzerland}
\affiliation{(4) CNRS, ENS de Lyon, LPENSL, UMR5672, 69342, Lyon cedex 07, France}
\preprint{APS/123-QED}

\title{Dynamical frustration in spacetime metamaterials enables cascading logic and synchronization}

\begin{abstract}
Spacetime metamaterials are engineered media whose constitutive parameters such as  permittivity, permeability, stiffness, or mass density are modulated simultaneously in both space and time. These additional degrees of freedom, absent in conventional static metamaterials, unlock a cabinet of wave phenomena that cannot be achieved in time-invariant structures, e.g. compact nonreciprocal devices, topological insulators, and devices for efficient frequency conversion and mixing and pulse shaping. The vast majority of these studies, however, operate in the stable regime, where modulation parameters are chosen to yield linear wave propagation. Here, we push spacetime metamaterials into the regime of parametric instability, and discover a novel type of ``dynamically frustrated'' oscillating states, where nonlinear non-reciprocal, topologically protected phase dislocations emerge. 
We control these dislocations and make them stop, split, and recombine. We harness this control to create devices for cascading logic in branched networks, and synchronization in 2D metamaterials.
Our findings are broadly applicable anywhere where spacetime modulation can be pushed beyond linear stability, from cold atoms and superconducting circuits to acoustics and RF circuits.

\end{abstract}
\maketitle
In the playground, every child swinging by themselves is enjoying a spontaneous $\mathbb{Z}_2$ symmetry breaking. When the swing is finally set in motion, it has picked up one of the two possible sub-harmonic limit cycles with a $0$ or $\pi$ phase. Such parametric oscillators can be found in scientists' playgrounds alike, e.g., water tanks \cite{douady_experimental_1990}, optical \cite{akhmanov_observation_1965,okawachi_demonstration_2020,marandi_network_2014} or mechanical \cite{carr_parametric_2000,grandi_enhancing_2021,mahboob_electromechanical_2016,mestre_network_2025} resonators, cold atoms~\cite{kim_quantum_2010,struck_quantum_2011}, optomechanical devices~\cite{braginsky_parametric_2001,del_pino_non-hermitian_2022,frimmer_rapid_2019}, and electrical circuits \cite{turner_five_1998,chou_analog_2019,cen_large-scale_2022}. Oftentimes, parametric oscillators are coupled to make their phase state interact, and are used to compute the ground state of Ising models
~\cite{heugel_ising_2022,bohm_poor_2019,mcmahon_fully_2016,utsunomiya_mapping_2011,kim_quantum_2010,struck_quantum_2011,cen_large-scale_2022,wang_coherent_2013,goto_parametron_1959,frimmer_rapid_2019,bashar_experimental_2020,mohseni_ising_2022,lucas_ising_2014}. 
Yet, the spatially homogeneous modulation of these machines strongly constrains their dynamics and the possible computing architectures.

\begin{figure}[b!]
    \centering
    \includegraphics[width=7.5cm]{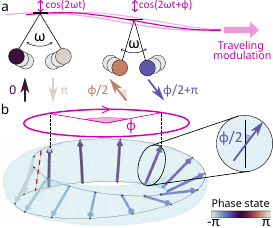}
    \caption{Dynamical frustration in spacetime media. (a) Chain of pendulums parametrically pumped by a traveling wave at $2\omega$. Each oscillator exhibits two limit cycles  with opposite phase $\phi/2$ and $\phi/2+\pi$, where $\phi$ is the local forcing phase imposed by the wave. (b) Closed chain with forcing phase winded from 0 to $2\pi$. Coupling tends to minimize phase differences between neighbors, such that the phase of the parametric oscillations winds from $0$ to $\pi$. Therefore, a  frustrated state spontaneously emerges in the form of a phase dislocation (red line). The phase states draw a M\"obius strip due to subharmonic response, showing topological protection of the frustration.
    }
    \label{fig:1}
\end{figure}

Here, we show that applying spacetime modulation to coupled parametric oscillators (Fig. \ref{fig:1}a) 
gives rise to a new class of nonlinear states, consisting of phase dislocations, that are non-reciprocal and topologically protected (Fig. \ref{fig:1}b). We then harness these ``dynamically frustrated'' states as information carriers for cascading logic \cite{liu_controlled_2024,kwakernaak_counting_2023} and synchronization.

Our findings sit at the crossroads of multiple areas of many-body physics, nonlinear physics and engineering---anywhere where non-reciprocity and nonlinearity interact. First, we push the field of spacetime metamaterials ~\cite{apffel_frequency_2022,wang_observation_2018,delory_elastic_2024,caloz_spacetime_2020,fleury_floquet_2016,yves_symmetry-driven_2026} beyond the stable regime, which has sofar been the dominant paradigm apart from few pioneering theoretical works~\cite{galiffi_broadband_2019,kruss_nondispersive_2022,melkani_space-time_2024,lambert_nonlinear_2026}.
Here we embrace nonlinearities instead and show that the phase dislocations, that are in fact non-reciprocal dark solitons, open avenues for wave guiding and computing, and could be further leveraged for sensing or robotic design~\cite{veenstra_adaptive_2025,veenstra_nonreciprocal_2025,slim_programmable_2025,belyansky_phase_2025}.
Second, our finding generalize the very concept of frustration. Geometric frustration stems from an incompatibility between local interactions and geometry~\cite{mellado_macroscopic_2012, wannier_antiferromagnetism_1950, drisko_topological_2017, balents_spin_2010, zhou_quantum_2017, wang_artificial_2006, pauling_structure_1935, struck_engineering_2013, kang_complex_2014, guo_non-orientable_2023, jorge_active_2024}. In stark contrast, dynamical frustration stems from an incompatibility between local dynamical states and geometry, which leads to unidirectional propagation and synchronization in 2D materials. This in turn enables robust cascading logic.

\begin{figure*}
    \centering
    \includegraphics[width=18cm]{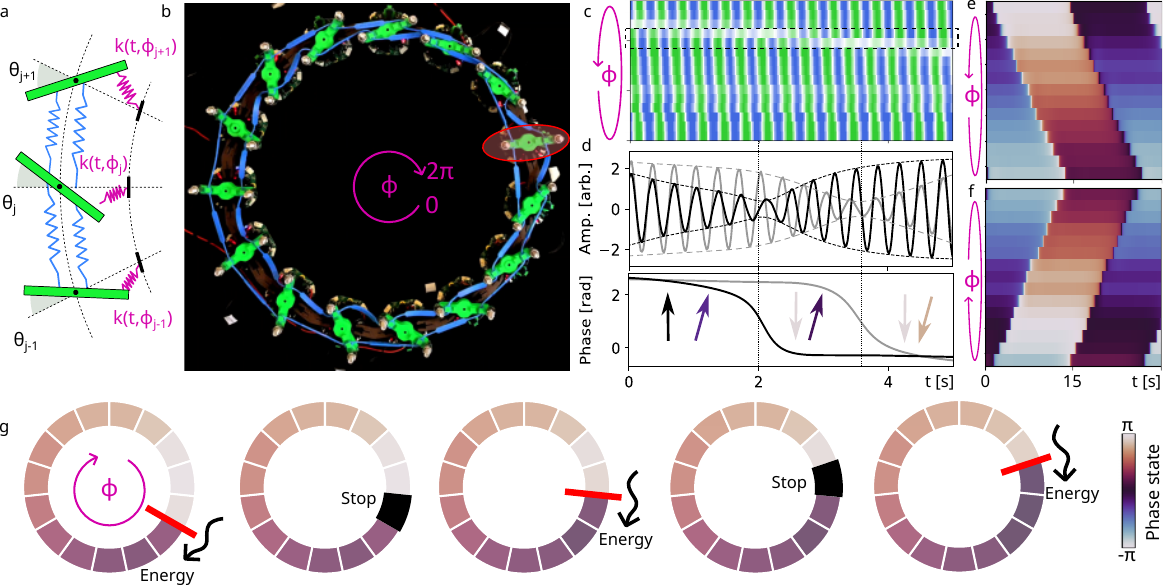}
    \caption{Non-reciprocal topologically protected phase dislocations. (a) Sketch of a piece of parametric chain. (b) Chain of coupled parametric oscillators displaying a phase dislocation. Green arrow shows winding of forcing phase. (c) Angular position of oscillators along time showing the propagation of the phase dislocation. (d) 
    A detail of such jump is shown, where both oscillator are initially in phase. The first one decreases its oscillation amplitude until (almost) cancellation at $t=57$ s, and starts again with opposite phase. The two oscillators remain out of phase until the second one performs similar phase switch at $t=58.5$ s. After the dislocation pass, both phase state have been switched by $\pi$ and oscillators are back in phase.
    (e) Kymograph of the slow phase state in the chain for anticlockwise and (f) clockwise winding. Dislocation propagation appears as a diagonal line.(g) Experimental measurement of phase state in the chain. Winding direction breaks the chiral symmetry and imposes unidirectional energy flux across the dislocation. Successive stops and restart of oscillators in opposite phase state allows the dislocation to travel against the phase gradient (see text for details).}
    \label{fig:2}
\end{figure*}
\section{Dynamical frustration}
Our unit cell is a parametric oscillator pumped at twice its resonant frequency with a tunable phase $\phi_j$. For large enough pumping, such oscillator reaches a subharmonic limit cycle with fixed phase $\phi_j/2$ or $\phi_j/2+\pi$ (Fig. \ref{fig:1}b). Coupling elastically $N=15$ unit cells placed along a closed loop makes up a 1D parametric metamaterial (Fig. \ref{fig:2}a-b). It is turned spacetime by winding the local forcing phase $\phi_j = 2Wj\pi/N$ between $0$ and $2W\pi$ along the chain, with $W \in \mathbb Z$. First choosing $W=1$, we observe upon visual inspection of the dynamics (Fig. \ref{fig:2}c and Supplementary Video 1) the emergence of a phase defect that forms a boundary between two domains oscillating in opposite phases. Strikingly, this defect moves unidirectionally. In fact, as the defect passes through each oscillator, it makes it momentarily slow down, stand still, and start again with the opposite phase (Fig. \ref{fig:2}d). 

We witness here the emergence of a non-reciprocal dynamical state in the form of a motile phase defect. Its existence is topologically enforced by the half-winding of the parametric oscillations $\phi_j/2\ [\pi]$ when the forcing phase $\phi_j$ performs a full winding. After a full turn of the dislocation, all oscillators changed their phase by $\pi$, and they all come back to their initial phase state after a second turn.
Like in geometric frustration, the emergent order parameter lives on a non-orientable manifold~\cite{guo_non-orientable_2023} (Fig.~1b). It has to travel twice about a M\"obius strip to return to the same point (Fig. \ref{fig:2}e). However, the analogy stops here. Geometric frustration originates from an unhappy balance between geometric constraints and leads to a myriad of degenerate ground states. In contrast, dynamic frustration leads to a single  unidirectional, oscillating state, which is controlled by the winding direction. In fact, the defect moves opposite to the winding (Fig. \ref{fig:2}e-f).

We now aim to provide a model for the defect's motion. First, we write the equations of motion describing each degree of freedom $\theta_j$ of our spacetime metamaterial 
\begin{equation}
\begin{split}
        \partial_t^2 \theta_j +  &2\Gamma \omega \partial_t\theta_j + \omega^2[1 + P\cos(2\omega t + \phi_j)] \theta_j = \\ 
        &-  \alpha(\theta_j) \theta_j+k (\theta_{j+1} + \theta_{j-1} - \theta_{j}) 
        \label{eq:coupledLinear}
\end{split}
\end{equation}
where $\omega=25$ rad.s$^{-1}$ is the resonance pulsation of a single oscillator, $\Gamma = 0.12$ the damping, $\alpha(\theta_j)$ the non-linear restoring onsite stiffness, $\sqrt k=8 $ rad.s$^{-1}$ the pulsation of the coupling springs (see SI for details on calibration). Parameters $P$ and $\phi_j$ are the tunable strength and local phase of the parametric modulation. 
Since we observe that the envelope of each oscillator varies slowly with respect to the carrier wave (Fig. \ref{fig:2}e), we perform a multiple scale expansion.
One seeks for solutions of Eq.~\eqref{eq:coupledLinear} of the form $\theta_j(t) = A_j(t)e^{i(\omega_0 t + \phi_j/2)} + h.c.$, where $A_j(t)$ is a slowly varying envelope that obeys the dimensionless equation (see SI or \cite{lambert_nonlinear_2026})
\begin{equation}
\begin{split}
       -i\dot A_j =&  pA_j^* +i\gamma A_j - \frac{3\tilde \alpha}{2}|A_j|^2A_j \\
       &-\kappa( e^{iW\pi/N}A_{j+1}+ e^{-iW\pi/N}A_{i-1} - 2 A_j )
       \label{eq:env}
\end{split}
\end{equation}
where $p, \gamma,\kappa, \tilde \alpha$ are the dimensionless forcing, damping coupling and nonlinearities. The $pA_j^*$ term is the energy injection by the pump, while coupling and phase winding lead to non-reciprocity $e^{\pm iW\pi/N}A_{j\pm1}$. It is reminiscent of synthetic gauge field in many-body physics~\cite{slim_programmable_2025,mancini_observation_2015,mathew_synthetic_2020,miyake_realizing_2013}, generated here by the forcing phase gradient. Doing so, one effectively implements non-reciprocal interactions already discussed for two- or three body parametric oscillators~\cite{calvanese_strinati_coherent_2020,bello_persistent_2019}.

Starting from small initial conditions in  Eq. (\ref{eq:env}), the most unstable linear mode of the chain admits a unique motile phase dislocation propagating at constant velocity (See SI). The exponential growth of such mode is eventually saturated by nonlinearities within a couple of seconds. Therefore, we now assume that all oscillators have reached a limit cycle of amplitude $\pm \theta_L$, such that the energy injected by the pump is compensated by nonlinearities (see SI). If the dislocation sits between site $j$ and $j+1$ (Fig. \ref{fig:2}g), the local energy $E_k=\langle \dot \theta_k^2\rangle$ averaged over one oscillation period around it evolves as
\begin{equation}
     \partial_t E_{j+1}=-\partial_t E_j =  2\Re(\theta_LA_j)/\tau \\
\end{equation}
with time scale for the amplitude decrease
\begin{equation}
    \tau = \frac{\omega}{\sin(W\pi/N) \omega_k^2}
    \label{eq:tau}
\end{equation} 
while $\partial_t E_k=0$ everywhere else in the chain. A non-zero flux appears only across the dislocation due to non-reciprocity. For positive winding, the energy flows from site $j$ to $j+1$ (Fig. \ref{fig:2}g). 

Therefore, oscillator $j$ will eventually stop ($t=2$ s in Fig. \ref{fig:2}d). The local pump can start it again in one of its two phase state, but the only way for this oscillator to not stop again is to go in phase with oscillator $j+1$. 
As such, it is now in phase opposition with oscillator $j-1$. Hence, the dislocation has leapfrogged backwards and the energy flows from site $j-1$ to $j$. This non-reciprocal energy transfer across the dislocation repeats cyclically and prescribes the dislocation velocity $v_0\sim1/\tau \sim 1.5$ site/s. Experimentally, we measure velocity of about 1 site/s, which is of the right order of magnitude. We checked that the dislocation velocity increases according to Eq.~(\ref{eq:tau}) with respect to the winding number $W$ (see SI), confirming that a larger phase difference between neighbors results in a faster energy transfer.
\section{Soliton and SNIC bifurcation}
So far, dislocation motion was studied from a 'local transfer' perspective. Nevertheless, its propagation is reminiscent of topological solitons in pendulum chains \cite{dauxois_physics_2010,frenkel_theory_1939}. To pursue such analogy, we approximate Eq.~\eqref{eq:env} as a non-reciprocal, driven version of the Ablowitz–Ladik equation, which is integrable (See SI). 
This equation admits nonlinear dark soliton solutions that describe the phase defects as
\begin{equation}
A_j(t) \propto\tanh(B(j-x(t)),
\label{eq:theoryProfile}
\end{equation}
where $B$ is a constant, and $x(t)$ the position of the center of mass of the soliton that obeys
\begin{equation}\label{eq:snic}
\frac{dx}{dt}\approx\frac{1}{\tau}[1+f(p)\cos (2\pi x) ],
\end{equation}
where $f(p)$ is a function given in SI, independent of $W$ in the limit $W/N \ll 1$. Thus, the velocity of the dislocation is predicted to depend on the amplitude. For pump levels close from the parametric instability threshold, $f(p)\approx 0$, the dark soliton is simply driven at a velocity $v_0=1/\tau$, in line with our previous analysis. For large enough forcing, $|f(p)|\geq 1$, the system develops fixed points near each site of the chain, by proceeding through a Saddle-Node on an Invariant Circle (SNIC) bifurcation~\cite{strogatz_nonlinear_2015}. This occurs for $p>p_c \sim 4 \sqrt{ \kappa^2 +\Gamma^2}$. Above this threshold, the dislocation is trapped by the closest site ($dx/dt=0$).

\begin{figure}
    \centering
    \includegraphics[width=8.5cm]{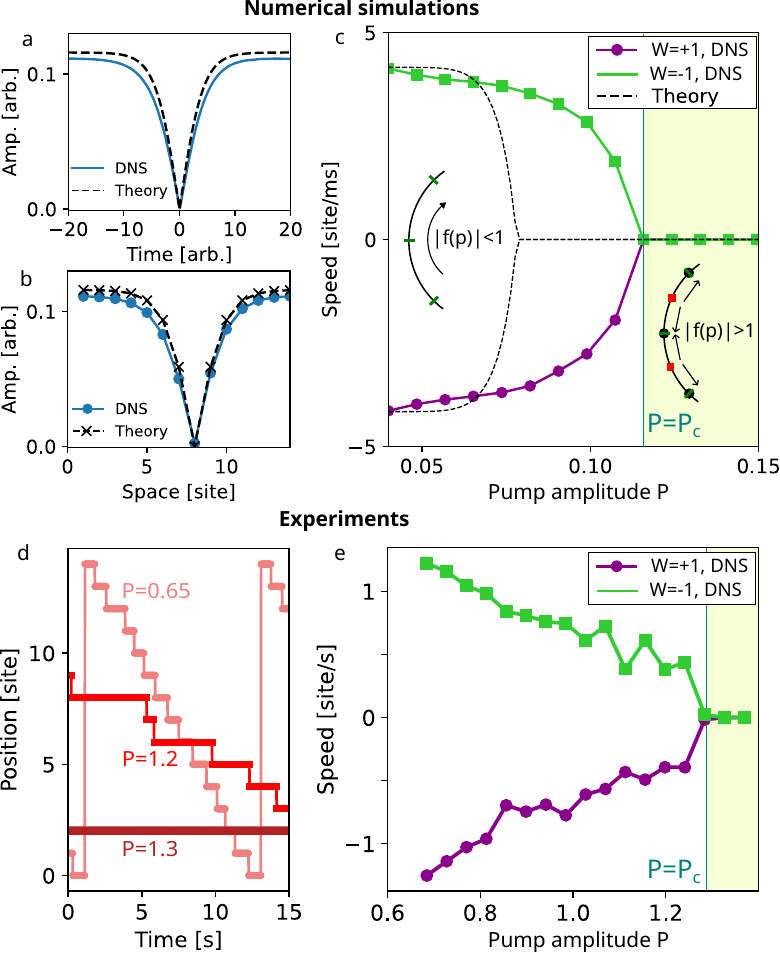}
    \caption{Non-reciprocal dark soliton undergoing a SNIC bifurcation. (a)  Envelope as a function of time and (b) space from direct numerical simulations (DNS) and soliton theory with no adjustable parameter (dashed line). (c) Velocity of the dislocation for $W=\pm1$ as a function of forcing amplitude $P$ from DNS (circles/squares) and theory (dashed line). Dislocation freezes for $P>P_c$. Repealing points  (red squares) appear for large enough forcing around each site (green dashes). (d) Experimental tracking of the dislocation for forcing amplitude of $P=0.65,1.2$ and $1.3$. Propagation goes from very regular to complete stopping. (e) Experimental result for velocity as function of forcing amplitude for $W=\pm1$. We do not obtain quantitative agreement for the critical amplitude $\epsilon_c$ in comparison to panel (c), presumably because the model does not capture viscoelastic dissipation or other nonlinear effects, as well as because forcing $\epsilon$ is not small compared to onsite potential.  }
    \label{fig:3}
\end{figure}

To validate this analysis, we first perform direct numerical simulations (DNS) of Eq. \eqref{eq:coupledLinear}, and compare it with the predicted profile (Eq. \eqref{eq:theoryProfile}) in Fig. \ref{fig:3}a-b. Both are in excellent agreement without any fitting parameter, whether one considers temporal (Fig. \ref{fig:3}a) or spatial (Fig. \ref{fig:3}b) profiles. The model also correctly captures the dependence of the velocity on the pump amplitude (Fig. \ref{fig:3}b). At lowest forcing, the predicted velocity obtained from numerical integration of Eq.~\eqref{eq:snic} (black dash) and the observed one (squares/circles) are in excellent agreement. On the other hand, the dislocation stops indeed for $p \sim p_c=0.07$. We attribute the quantitative mismatch between the predicted and observed critical forcing to all the approximations performed along the way to the soliton description. Further comparison between DNS and theory shown in SI confirm that the SNIC bifurcation picture captures all the observed physics.

Such transition is also present in our experiments. As the pumping amplitude $P$ is increased, the dark soliton takes longer steps and eventually stops (Fig. \ref{fig:3}d). This bifurcation is reflected by the average velocity, which progressively decreases with the pumping $P$ and reaches zero at a critical value $P_c$ (Fig. \ref{fig:3}e). Due to coupling inhomogeneities, some sites reach their SNIC bifurcation at lower forcing amplitude than others. This explains why the soliton takes longer steps on some sites, as well as why it stops at preferred locations close from the bifurcation. Upon further increasing forcing, the dislocation can get trapped by any site, as all of them went through their bifurcation.

Before going further, we emphasize two important comments on the soliton analysis. First, the motion of the dark soliton is not phase-locked with the fast external forcing. Hence, the slowly varying collective limit cycle performs a continuous time-symmetry breaking, in strong contrast with local phase states that perform discrete symmetry breaking. This is reminiscent to what happens in continuous (space-)time crystals \cite{zhao_space-time_2025,xu_space-time_2018,smits_observation_2018,kongkhambut_observation_2022,veenstra_wave_2025}. Second, topological solitons generically connect locally stable states, whose multiplicity is rooted in nonlinearity~\cite{dauxois_physics_2010}. In our system, multistability arises from phase locking of subharmonic parametric oscillators, which is a linear process. Hence, the present phase dislocation still exists and moves unidirectionally, even in the absence of nonlinearities (see SI). Nevertheless, the existence of nonlinearities in our system makes the soliton analysis suitable, and strongly enriches the physics as the SNIC bifurcation scenario is a purely nonlinear process. 

\section{Soliton source and computing}

\begin{figure}
    \centering
    \includegraphics[width=8.5cm]{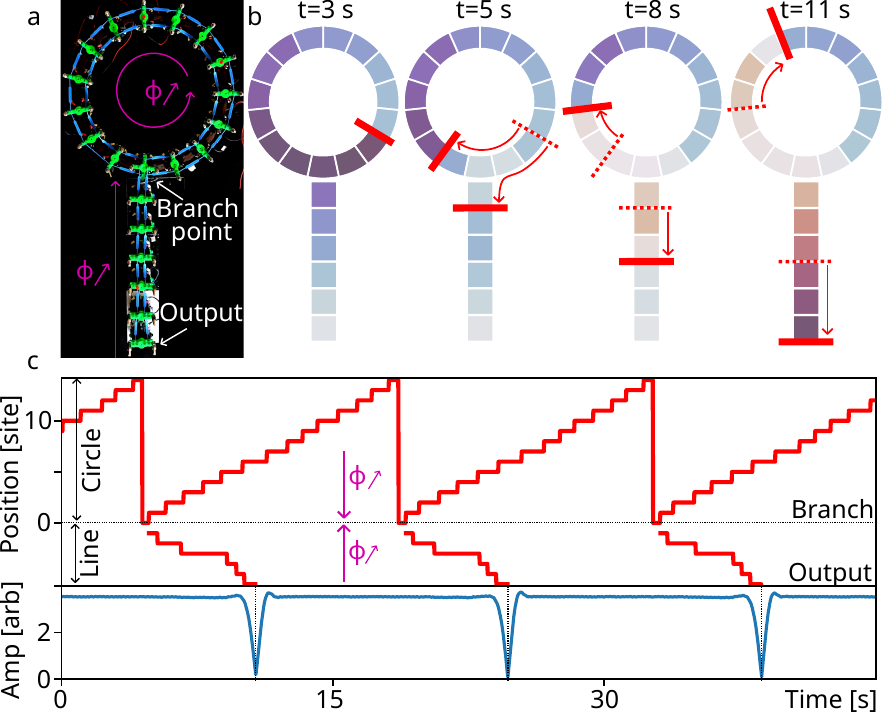}
    \caption{Controlled source of dark solitons. (a) Picture of a transmission line coupled to a circular chain. Phase gradient in the line enforces unidirectional propagation toward the output. (b) Experimental measurement of phase states at various times. When the frustration goes through the crossing point, it splits in two. The frustration in the line is evacuated through the open end, the one in circle keeps propagating. (c) Trajectories of frustrations and oscillation amplitude of the last oscillator of the line. The device acts a periodic source of dark solitons driven by the circular motion of the frustration. Each time the ring's soliton hits the branching point, a soliton is unidirectionally transmitted to the free output of the line. The slow oscillation amplitude of the output oscillator (lower panel) is nearly constant, except on very short time windows where it drops toward zero due to the dark soliton's arrival.}
    \label{fig:4}
\end{figure}

Our setup consistently produces dark solitons that remain trapped on a ring. In Fig. \ref{fig:4}, we untrap solitons through an ``exit door", which turns our system into a periodic source of dark solitons. We do so by mechanically coupling the circle with a line-chain, which phase gradient is directed toward the circle (Fig. \ref{fig:4}a).  When the pump is turned on, a dark soliton appears in the circle and travels clockwise as shown in Fig. \ref{fig:4}b ($t=3$ s). At $t=5$ s, the dislocation travels through the branching point and switches its phase state. As a consequence, the line and the circle now exhibit a phase dislocation. Both dislocations keep travelling against the phase gradient ($t=8$ s). The  dislocation in the line is eventually evacuated towards the open end ($t= 11$ s), which completes the line's phase switch. The dislocation in the ring keeps propagating, and the same phenomenon repeats once the dislocation has performed a full turn (Fig. \ref{fig:4}c). 
Consequently, the slow oscillation amplitude of the output oscillator (lower panel) is nearly constant, except on very short time windows where it drops toward zero due to the dark soliton's arrival. 
Thanks to the generality of our formalism, the proposed design for a dark soliton source can be directly transposed to any other parametric system, including optics \cite{mcmahon_fully_2016}, microwaves \cite{nasari_observation_2026} or cold atoms \cite{smits_observation_2018}.

\begin{figure*}
    \centering
    \includegraphics[width=17cm]{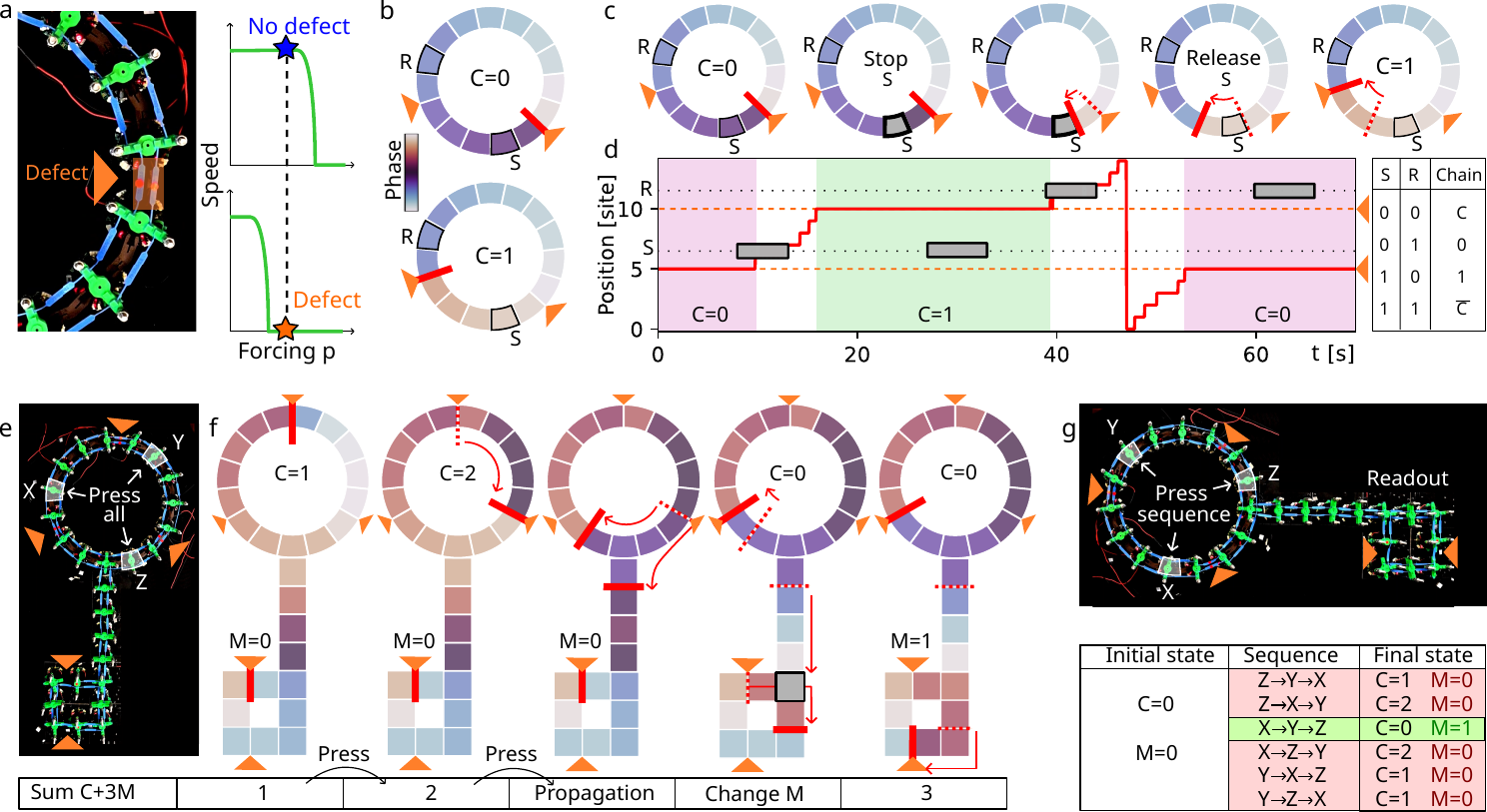}
     \caption{Cascading logic enabled by dynamical frustration in spacetime branched networks. (a) Picture of a coupling defect (orange triangle). Forcing is chosen such that the SNIC bifurcation occurs at the defect but not on the other sites, which blocks the frustration on the defect site. (b) A chain with two defects behaves as a mechanical memory. The memory states $C=0,1$ correspond to frustration trapped at different locations. The two input pins S and R serve to control the memory state. (c) Example of state change from $C=0$ to $C=1$. If one stops the pin $S$, the dislocation initially on $C=0$ moves toward S. Now releasing $S$, the dislocation propagates clockwise and stops at the next defect, which corresponds to $C=1$ (d) Trajectory of the frustration along time when various press/release events (gray rectangles) are triggered. Trying all possible combinations and initial states, one obtains the truth table of a JK flip-flop. (e) Picture of a sequential counter. A circular chain with 3 defects (states $C=0,1,2$) is connected to a square chain with 2 defects (states $M=0,1$). Frustration in the line can only propagate from the circle to the square. A counting event corresponds to press/release X,Y,Z simultaneously. (f) Experimental measurement of phase states after each counting event. When the circles goes from $C=2$ to $C=0$, a frustration travels through the line and induces a transition in the square from $M=0$ to $M=1$. (g) The same device, initially in $C=0, M=0$ can also be used as a sequence detector. Upon pressing a 3 letter combination on the circle (=keyboard), only the sequence $X \rightarrow Y \rightarrow Z$ changes the state of the square (=readout) from 0 to 1.}
    \label{fig:5}
\end{figure*}

The previous setup also offers a way to perform unidirectional transport of information that we now exploit to create devices for cascading logic. First, we take advantage of the SNIC bifurcation discussed above to create stable soliton positions, that will act as memory states. We do so by engineering coupling defects in the chain, shown Fig. \ref{fig:5}a, such that soliton only blocks at the defects where the SNIC-bifurcation has been crossed. Setting two defects, one obtains a chain with two possible dislocation locations denoted $C=0,1$  (Fig. \ref{fig:5}b). 

The ring state can be changed upon acting on the clockwise neighbor of the defects, called `input pins', denoted S and R (Fig. \ref{fig:5}b). A `pin activation' is defined as an external stop of its motion for a short amount of time, which is done by pressing the oscillator for a couple of seconds with fingers before releasing it. 
If the ring is initially in state $C=0$ (Fig. \ref{fig:5}c-d, and Supplementary Video XXX), 
when S is activated, the soliton jumps on it. When S is released, the soliton does not go backward to $C=0$, but instead moves clockwise until it is trapped by the next defect $C=1$. 
For instance, activating $S$ sets the chain in state $C=1$ whatever the initial state, while activating R always resets the chain in $C=0$. We measured experimentally the full truth table of our device, and found that our chain behaves as a mechanical JK flip-flop.

We now couple several mechanical memories to perform complex sequential computing \cite{liu_controlled_2024,kwakernaak_counting_2023,guo_non-orientable_2023} . For instance, a counter where a first circular chain $C=0,1,2$ encodes a first digit and a second square chain $M=0,1$ encodes potential carry due to overflow of $C$, is shown in Fig. \ref{fig:5}e. A transmission line, as in Fig. \ref{fig:4}, connects the section of the circle between $C=2$ and $C=0$, and the input pin of the square near $M=0$ (Fig. \ref{fig:5}a-b). A counting event is defined as a simultaneous activation of the three circle's pins $X,Y,Z$, and the total count is encoded in $S=C+3M$. Starting from $(C=1,M=0)$, a first counting event gives $(C=2,M=0)$. When a second counting event occurs, the dislocation in the circle hits the branching point with the line and splits in two (Fig. \ref{fig:5}f). The dislocation in the circle quickly reaches the next state $C=0$. The dark soliton in the line propagates, reaches the square cell, which activates (i.e. stops) the input pin of $M=0$. This activation triggers a transition toward the state $M=1$. Overall, the counter kept track of the overflow  $C=2 \rightarrow 0$ and saved the carry in the square state $M=0 \rightarrow 1$.

The same device also allows sequence detection. Consider now the circle as a keyboard $\{X,Y,Z\}$, and the square as a readout (Fig. \ref{fig:5}g). Starting from $(C=0,M=0)$, only the sequence $X\rightarrow Y \rightarrow Z$ generates a change in the square state 
$M=0\rightarrow1$. Indeed, it is the only sequence for which the dislocation in the circle performs a full revolution, which cascades a dark soliton toward the readout. 

The proposed design exhibits desirable properties for larger cascading logic. The ability to split dislocations at a circuit node allows to cascade information between different memories, while unidirectional transmission clearly sets the a hierarchy between a parent process (the circle) and a worker (the square). Hence, scalability appears much easier than in previous pioneering works on mechanical computers, that used physical inhomogeneities to introduce hierarchy \cite{liu_controlled_2024,kwakernaak_counting_2023}.  Moreover, the active nature of the material prevents dissipation to limit information propagation upon up-scaling. Last, the generality of the formalism makes it transposable to any experimental platform with parametric oscillators. 

\section{Degeneracy and synchronization}

\begin{figure*}
    \centering
    \includegraphics[width=17cm]{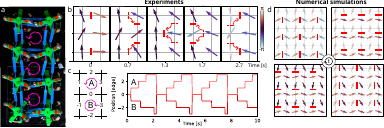}
     \caption{Frustration in a two-dimensional network. (a) Tesselation of chains with opposite winding using X-shaped pendulums. (b) Phase state of two adjacent cells at different times. Red line materializes the dislocations, which are synchronized (see Supplementary Movie for the full metamaterial motion). (c) Location of both dislocations along time. Regular visit of frustration on the shared edge triggers synchronization. (d) Numerical simulation on 6x6 network with fixed boundary condition, showing global synchronization of the network.}
    \label{fig:6}
\end{figure*}

Another strong aspect of dynamical frustration is its effect on phase state exploration. Indeed, there exists $2N=30$ possible chain states that display a single dislocation. During its propagation, the dislocation flips all the phase states one after another (Fig. \ref{fig:2}g), and all the $30$ possible chain states are visited in an ordered manner. This provides a natural connection between geometrical and dynamical frustration: dynamically frustrated states explores all the possible static ground states. However, this only holds for 1D system, as a completely different phenomenology emerges in a 2D dynamically frustrated systems. These are obtained by tessellating chains with opposite winding on a square lattice as in Fig. \ref{fig:6}a, and realized experimentally using X-shaped pendulums elastically coupled to 4 neighbors. 

As before, frustration propagates locally with direction opposite to winding, and each cell explores all its ground states. However, the dynamic of neighboring cell is not independent, as as shown in Fig. \ref{fig:6}b. Indeed, these must share their dislocation at some time through their common edge. As a consequence, the dynamic of neighboring cells is strongly correlated, and phase state are spatially symmetrical with respect to the share edges at (almost) all times. It is confirmed by Fig. \ref{fig:6}c, where the position of the dislocations along time is shown and display clear splitting and recombination events. These splitting events are similar to those observed in Fig. \ref{fig:4}, but the line is now connected through its both ends to the circle, which triggers the synchronization.

Hence, dynamical frustration induces local synchronization, and the whole network eventually beats in phase (see also Supplementary video). Thus, the system does not explore all the possible network states with a single dislocation per cell, but only a marginal part of it.  The process can be reproduced numerically on larger networks, as shown in Fig. \ref{fig:6}d for 6x6 system.
Whatever the size of the material, fixing the location of a single frustration at a single time in the network is enough to determine the location of all frustrations at all time. Thus, the number of explored states is not extensive with the system size, but rather fixed by the number of states of a single cell. This stands in strong contrast with 2D geometric frustration, where the number of explored states grows with the system size \cite{pauling_structure_1935}. 

Our results open several interesting perspectives. First, the new type of chiral dynamical state augments the toolkit of temporal \cite{wilczek_quantum_2012,bruno_comment_2013,zhang_observation_2017} and spatio-temporal \cite{smits_observation_2018,xu_space-time_2018,zhao_space-time_2025,melkani_space-time_2024} crystals. The proposed mechanism for recovering continuous temporal symmetry at large time scales despite discrete symmetry breaking at fast time scale may be extended to other systems. Sensitivity of the proposed 2d-synchronization with respect to noise. Last, our method also provides a route toward engineering of self-appearing chiral nonlinear limit cycles, which may have, on top of the already realized cascading logic, implications in robotics, active matter \cite{brandenbourger_non-reciprocal_2019,pedergnana_loss-compensated_2024}, analog computing in coherent Ising machines~\cite{calvanese_strinati_coherent_2020, cen_phase-diagram_2023,zhou_frustration_2025}, and physical learning~\cite{de_bos_multimodal_2026,du_metamaterials_2026}.
%

\emph{Acknowledgments---}The authors would like to thank Daan Giesen for technical assistance. R.M. and C.C. acknowledge funding from the Netherlands Organisation for Scientific Research (Grant Agreement No. VIDI 2131313). C.C. acknowledges funding from the European Research Council (Grant Agreement ERC-CoG 101170693).

\emph{Data availability statement---all the data supporting this study are publicly available at XXXXXX.}

\bibliographystyle{naturemag}
\bibliography{biblio}

@article{lucas_ising_2014,
	title = {Ising formulations of many {NP} problems},
	volume = {Volume 2 - 2014},
	issn = {2296-424X},
	url = {https://www.frontiersin.org/journals/physics/articles/10.3389/fphy.2014.00005},
	doi = {10.3389/fphy.2014.00005},
	journal = {Frontiers in Physics},
	author = {Lucas, Andrew},
	year = {2014},
}

@article{utsunomiya_mapping_2011,
	title = {Mapping of {Ising} models onto injection-locked laser systems},
	volume = {19},
	copyright = {\&\#169; 2011 OSA},
	issn = {1094-4087},
	url = {https://opg.optica.org/oe/abstract.cfm?uri=oe-19-19-18091},
	doi = {10.1364/OE.19.018091},
	language = {EN},
	number = {19},
	urldate = {2022-11-02},
	journal = {Optics Express},
	author = {Utsunomiya, Shoko and Takata, Kenta and Yamamoto, Yoshihisa},
	month = sep,
	year = {2011},
	pages = {18091--18108},
}

@article{frimmer_rapid_2019,
	title = {Rapid {Flipping} of {Parametric} {Phase} {States}},
	volume = {123},
	issn = {0031-9007, 1079-7114},
	url = {https://link.aps.org/doi/10.1103/PhysRevLett.123.254102},
	doi = {10.1103/PhysRevLett.123.254102},
	language = {en},
	number = {25},
	urldate = {2023-01-23},
	journal = {Physical Review Letters},
	author = {Frimmer, Martin and Heugel, Toni L. and Nosan, Žiga and Tebbenjohanns, Felix and Hälg, David and Akin, Abdulkadir and Degen, Christian L. and Novotny, Lukas and Chitra, R. and Zilberberg, Oded and Eichler, Alexander},
	month = dec,
	year = {2019},
	pages = {254102},
}

@article{goto_parametron_1959,
	title = {The {Parametron}, a {Digital} {Computing} {Element} {Which} {Utilizes} {Parametric} {Oscillation}},
	volume = {47},
	issn = {2162-6634},
	doi = {10.1109/JRPROC.1959.287195},
	number = {8},
	journal = {Proceedings of the IRE},
	author = {Goto, Eiichi},
	month = aug,
	year = {1959},
	pages = {1304--1316},
}

@article{turner_five_1998,
	title = {Five parametric resonances in a microelectromechanical system},
	volume = {396},
	copyright = {1998 Macmillan Magazines Ltd.},
	issn = {1476-4687},
	url = {https://www.nature.com/articles/24122},
	doi = {10.1038/24122},
	language = {en},
	number = {6707},
	urldate = {2023-08-07},
	journal = {Nature},
	author = {Turner, Kimberly L. and Miller, Scott A. and Hartwell, Peter G. and MacDonald, Noel C. and Strogatz, Steven H. and Adams, Scott G.},
	month = nov,
	year = {1998},
	pages = {149--152},
}

@article{douady_experimental_1990,
	title = {Experimental study of the {Faraday} instability},
	volume = {221},
	issn = {1469-7645, 0022-1120},
	url = {https://www.cambridge.org/core/journals/journal-of-fluid-mechanics/article/abs/experimental-study-of-the-faraday-instability/DB294255CBCDF217BFA024192067E861},
	doi = {10.1017/S0022112090003603},
	language = {en},
	urldate = {2023-08-08},
	journal = {Journal of Fluid Mechanics},
	author = {Douady, S.},
	month = dec,
	year = {1990},
	pages = {383--409},
}

@article{carr_parametric_2000,
	title = {Parametric amplification in a torsional microresonator},
	volume = {77},
	issn = {0003-6951, 1077-3118},
	url = {https://pubs.aip.org/apl/article/77/10/1545/515762/Parametric-amplification-in-a-torsional},
	doi = {10.1063/1.1308270},
	language = {en},
	number = {10},
	urldate = {2023-08-31},
	journal = {Applied Physics Letters},
	author = {Carr, Dustin W. and Evoy, Stephane and Sekaric, Lidija and Craighead, H. G. and Parpia, J. M.},
	month = sep,
	year = {2000},
	pages = {1545--1547},
}

@article{akhmanov_observation_1965,
	title = {Observation of {Parametric} {Amplification} in the {Optical} {Range}},
	volume = {2},
	journal = {Jetp Letters - JETP LETT-ENGL TR},
	author = {Akhmanov, S. and Kovrigin, A. and Piskarskas, Algis and Fadeev, V. and Khokhlov, R.},
	month = jan,
	year = {1965},
}

@article{grandi_enhancing_2021,
	title = {Enhancing and controlling parametric instabilities in mechanical systems},
	volume = {43},
	issn = {2352-4316},
	url = {https://www.sciencedirect.com/science/article/pii/S2352431621000195},
	doi = {10.1016/j.eml.2021.101195},
	urldate = {2023-08-31},
	journal = {Extreme Mechanics Letters},
	author = {Grandi, Alvaro A. and Protière, Suzie and Lazarus, Arnaud},
	month = feb,
	year = {2021},
	pages = {101195},
}

@article{zhou_frustration_2025,
	title = {Frustration {Elimination} and {Excited} {State} {Search} in {Coherent} {Ising} {Machines}},
	volume = {134},
	url = {https://link.aps.org/doi/10.1103/PhysRevLett.134.090401},
	doi = {10.1103/PhysRevLett.134.090401},
	number = {9},
	urldate = {2025-04-16},
	journal = {Physical Review Letters},
	author = {Zhou, Zheng-Yang and Gneiting, Clemens and You, J. Q. and Nori, Franco},
	month = mar,
	year = {2025},
	pages = {090401},
}

@article{mcmahon_fully_2016,
	title = {A fully programmable 100-spin coherent {Ising} machine with all-to-all connections},
	volume = {354},
	url = {https://www.science.org/doi/10.1126/science.aah5178},
	doi = {10.1126/science.aah5178},
	number = {6312},
	urldate = {2025-04-16},
	journal = {Science},
	author = {McMahon, Peter L. and Marandi, Alireza and Haribara, Yoshitaka and Hamerly, Ryan and Langrock, Carsten and Tamate, Shuhei and Inagaki, Takahiro and Takesue, Hiroki and Utsunomiya, Shoko and Aihara, Kazuyuki and Byer, Robert L. and Fejer, M. M. and Mabuchi, Hideo and Yamamoto, Yoshihisa},
	month = nov,
	year = {2016},
	pages = {614--617},
}

@article{cen_large-scale_2022,
	title = {Large-scale coherent {Ising} machine based on optoelectronic parametric oscillator},
	volume = {11},
	copyright = {2022 The Author(s)},
	issn = {2047-7538},
	url = {https://www.nature.com/articles/s41377-022-01013-1},
	doi = {10.1038/s41377-022-01013-1},
	language = {en},
	number = {1},
	urldate = {2025-04-16},
	journal = {Light: Science \& Applications},
	author = {Cen, Qizhuang and Ding, Hao and Hao, Tengfei and Guan, Shanhong and Qin, Zhiqiang and Lyu, Jiaming and Li, Wei and Zhu, Ninghua and Xu, Kun and Dai, Yitang and Li, Ming},
	month = nov,
	year = {2022},
	pages = {333},
}

@article{mohseni_ising_2022,
	title = {Ising machines as hardware solvers of combinatorial optimization problems},
	volume = {4},
	copyright = {2022 Springer Nature Limited},
	issn = {2522-5820},
	url = {https://www.nature.com/articles/s42254-022-00440-8},
	doi = {10.1038/s42254-022-00440-8},
	language = {en},
	number = {6},
	urldate = {2025-04-16},
	journal = {Nature Reviews Physics},
	author = {Mohseni, Naeimeh and McMahon, Peter L. and Byrnes, Tim},
	month = jun,
	year = {2022},
	pages = {363--379},
}

@article{marandi_network_2014,
	title = {Network of time-multiplexed optical parametric oscillators as a coherent {Ising} machine},
	volume = {8},
	copyright = {2014 Springer Nature Limited},
	issn = {1749-4893},
	url = {https://www.nature.com/articles/nphoton.2014.249},
	doi = {10.1038/nphoton.2014.249},
	language = {en},
	number = {12},
	urldate = {2025-04-24},
	journal = {Nature Photonics},
	author = {Marandi, Alireza and Wang, Zhe and Takata, Kenta and Byer, Robert L. and Yamamoto, Yoshihisa},
	month = dec,
	year = {2014},
	pages = {937--942},
}

@article{chou_analog_2019,
	title = {Analog {Coupled} {Oscillator} {Based} {Weighted} {Ising} {Machine}},
	volume = {9},
	copyright = {2019 The Author(s)},
	issn = {2045-2322},
	url = {https://www.nature.com/articles/s41598-019-49699-5},
	doi = {10.1038/s41598-019-49699-5},
	language = {en},
	number = {1},
	urldate = {2025-04-24},
	journal = {Scientific Reports},
	author = {Chou, Jeffrey and Bramhavar, Suraj and Ghosh, Siddhartha and Herzog, William},
	month = oct,
	year = {2019},
	pages = {14786},
}

@article{okawachi_demonstration_2020,
	title = {Demonstration of chip-based coupled degenerate optical parametric oscillators for realizing a nanophotonic spin-glass},
	volume = {11},
	copyright = {2020 The Author(s)},
	issn = {2041-1723},
	url = {https://www.nature.com/articles/s41467-020-17919-6},
	doi = {10.1038/s41467-020-17919-6},
	language = {en},
	number = {1},
	urldate = {2025-04-24},
	journal = {Nature Communications},
	author = {Okawachi, Yoshitomo and Yu, Mengjie and Jang, Jae K. and Ji, Xingchen and Zhao, Yun and Kim, Bok Young and Lipson, Michal and Gaeta, Alexander L.},
	month = aug,
	year = {2020},
	pages = {4119},
}

@article{struck_quantum_2011,
	title = {Quantum {Simulation} of {Frustrated} {Classical} {Magnetism} in {Triangular} {Optical} {Lattices}},
	volume = {333},
	url = {https://www.science.org/doi/10.1126/science.1207239},
	doi = {10.1126/science.1207239},
	number = {6045},
	urldate = {2025-04-24},
	journal = {Science},
	author = {Struck, J. and Ölschläger, C. and Le Targat, R. and Soltan-Panahi, P. and Eckardt, A. and Lewenstein, M. and Windpassinger, P. and Sengstock, K.},
	month = aug,
	year = {2011},
	pages = {996--999},
}

@article{bohm_poor_2019,
	title = {A poor man’s coherent {Ising} machine based on opto-electronic feedback systems for solving optimization problems},
	volume = {10},
	copyright = {2019 The Author(s)},
	issn = {2041-1723},
	url = {https://www.nature.com/articles/s41467-019-11484-3},
	doi = {10.1038/s41467-019-11484-3},
	language = {en},
	number = {1},
	urldate = {2025-04-24},
	journal = {Nature Communications},
	author = {Böhm, Fabian and Verschaffelt, Guy and Van der Sande, Guy},
	month = aug,
	year = {2019},
	pages = {3538},
}

@article{kim_quantum_2010,
	title = {Quantum simulation of frustrated {Ising} spins with trapped ions},
	volume = {465},
	copyright = {2010 Springer Nature Limited},
	issn = {1476-4687},
	url = {https://www.nature.com/articles/nature09071},
	doi = {10.1038/nature09071},
	language = {en},
	number = {7298},
	urldate = {2025-04-24},
	journal = {Nature},
	author = {Kim, K. and Chang, M.-S. and Korenblit, S. and Islam, R. and Edwards, E. E. and Freericks, J. K. and Lin, G.-D. and Duan, L.-M. and Monroe, C.},
	month = jun,
	year = {2010},
	pages = {590--593},
}

@article{heugel_ising_2022,
	title = {Ising machines with strong bilinear coupling},
	volume = {4},
	url = {https://link.aps.org/doi/10.1103/PhysRevResearch.4.013149},
	doi = {10.1103/PhysRevResearch.4.013149},
	number = {1},
	urldate = {2025-04-24},
	journal = {Physical Review Research},
	author = {Heugel, Toni L. and Zilberberg, Oded and Marty, Christian and Chitra, R. and Eichler, Alexander},
	month = feb,
	year = {2022},
	pages = {013149},
}

@article{bashar_experimental_2020,
	title = {Experimental {Demonstration} of a {Reconfigurable} {Coupled} {Oscillator} {Platform} to {Solve} the {Max}-{Cut} {Problem}},
	volume = {6},
	issn = {2329-9231},
	url = {https://ieeexplore.ieee.org/abstract/document/9204635},
	doi = {10.1109/JXCDC.2020.3025994},
	number = {2},
	urldate = {2025-04-24},
	journal = {IEEE Journal on Exploratory Solid-State Computational Devices and Circuits},
	author = {Bashar, Mohammad Khairul and Mallick, Antik and Truesdell, Daniel S. and Calhoun, Benton H. and Joshi, Siddharth and Shukla, Nikhil},
	month = dec,
	year = {2020},
	pages = {116--121},
}

@article{struck_engineering_2013,
	title = {Engineering {Ising}-{XY} spin-models in a triangular lattice using tunable artificial gauge fields},
	volume = {9},
	copyright = {2013 Springer Nature Limited},
	issn = {1745-2481},
	url = {https://www.nature.com/articles/nphys2750},
	doi = {10.1038/nphys2750},
	language = {en},
	number = {11},
	urldate = {2025-04-25},
	journal = {Nature Physics},
	author = {Struck, J. and Weinberg, M. and Ölschläger, C. and Windpassinger, P. and Simonet, J. and Sengstock, K. and Höppner, R. and Hauke, P. and Eckardt, A. and Lewenstein, M. and Mathey, L.},
	month = nov,
	year = {2013},
	pages = {738--743},
}

@article{drisko_topological_2017,
	title = {Topological frustration of artificial spin ice},
	volume = {8},
	copyright = {2017 The Author(s)},
	issn = {2041-1723},
	url = {https://www.nature.com/articles/ncomms14009},
	doi = {10.1038/ncomms14009},
	language = {en},
	number = {1},
	urldate = {2025-04-28},
	journal = {Nature Communications},
	author = {Drisko, Jasper and Marsh, Thomas and Cumings, John},
	month = jan,
	year = {2017},
	pages = {14009},
}

@article{zhou_quantum_2017,
	title = {Quantum spin liquid states},
	volume = {89},
	url = {https://link.aps.org/doi/10.1103/RevModPhys.89.025003},
	doi = {10.1103/RevModPhys.89.025003},
	number = {2},
	urldate = {2025-04-28},
	journal = {Reviews of Modern Physics},
	author = {Zhou, Yi and Kanoda, Kazushi and Ng, Tai-Kai},
	month = apr,
	year = {2017},
	pages = {025003},
}

@article{balents_spin_2010,
	title = {Spin liquids in frustrated magnets},
	volume = {464},
	copyright = {2010 Springer Nature Limited},
	issn = {1476-4687},
	url = {https://www.nature.com/articles/nature08917},
	doi = {10.1038/nature08917},
	language = {en},
	number = {7286},
	urldate = {2025-04-28},
	journal = {Nature},
	author = {Balents, Leon},
	month = mar,
	year = {2010},
	pages = {199--208},
}

@article{cen_phase-diagram_2023,
	title = {Phase-diagram investigation of frustrated {1D} and {2D} {Ising} models in {OEO}-based {Ising} machine},
	volume = {48},
	copyright = {© 2023 Optica Publishing Group},
	issn = {1539-4794},
	url = {https://opg.optica.org/ol/abstract.cfm?uri=ol-48-21-5459},
	doi = {10.1364/OL.499385},
	language = {EN},
	number = {21},
	urldate = {2025-04-28},
	journal = {Optics Letters},
	author = {Cen, Qizhuang and Ding, Hao and Guan, Shanhong and Hao, Tengfei and Li, Wei and Zhu, NingHua and Dai, Yitang and Li, Ming},
	month = nov,
	year = {2023},
	pages = {5459--5462},
}

@article{brandenbourger_non-reciprocal_2019,
	title = {Non-reciprocal robotic metamaterials},
	volume = {10},
	copyright = {2019 The Author(s)},
	issn = {2041-1723},
	url = {https://www.nature.com/articles/s41467-019-12599-3},
	doi = {10.1038/s41467-019-12599-3},
	language = {en},
	number = {1},
	urldate = {2025-04-28},
	journal = {Nature Communications},
	author = {Brandenbourger, Martin and Locsin, Xander and Lerner, Edan and Coulais, Corentin},
	month = oct,
	year = {2019},
	pages = {4608},
}

@article{wang_coherent_2013,
	title = {Coherent {Ising} machine based on degenerate optical parametric oscillators},
	volume = {88},
	url = {https://link.aps.org/doi/10.1103/PhysRevA.88.063853},
	doi = {10.1103/PhysRevA.88.063853},
	number = {6},
	urldate = {2025-04-28},
	journal = {Physical Review A},
	author = {Wang, Zhe and Marandi, Alireza and Wen, Kai and Byer, Robert L. and Yamamoto, Yoshihisa},
	month = dec,
	year = {2013},
	pages = {063853},
}

@article{mellado_macroscopic_2012,
	title = {Macroscopic {Magnetic} {Frustration}},
	volume = {109},
	url = {https://link.aps.org/doi/10.1103/PhysRevLett.109.257203},
	doi = {10.1103/PhysRevLett.109.257203},
	number = {25},
	urldate = {2025-05-15},
	journal = {Physical Review Letters},
	author = {Mellado, Paula and Concha, Andres and Mahadevan, L.},
	month = dec,
	year = {2012},
	pages = {257203},
}

@article{wang_artificial_2006,
	title = {Artificial ‘spin ice’ in a geometrically frustrated lattice of nanoscale ferromagnetic islands},
	volume = {439},
	copyright = {2006 Springer Nature Limited},
	issn = {1476-4687},
	url = {https://www.nature.com/articles/nature04447},
	doi = {10.1038/nature04447},
	language = {en},
	number = {7074},
	urldate = {2025-05-15},
	journal = {Nature},
	author = {Wang, R. F. and Nisoli, C. and Freitas, R. S. and Li, J. and McConville, W. and Cooley, B. J. and Lund, M. S. and Samarth, N. and Leighton, C. and Crespi, V. H. and Schiffer, P.},
	month = jan,
	year = {2006},
	pages = {303--306},
}

@article{pedergnana_loss-compensated_2024,
	title = {Loss-compensated non-reciprocal scattering based on synchronization},
	volume = {15},
	copyright = {2024 The Author(s)},
	issn = {2041-1723},
	url = {https://www.nature.com/articles/s41467-024-51373-y},
	doi = {10.1038/s41467-024-51373-y},
	language = {en},
	number = {1},
	urldate = {2025-05-15},
	journal = {Nature Communications},
	author = {Pedergnana, Tiemo and Faure-Beaulieu, Abel and Fleury, Romain and Noiray, Nicolas},
	month = aug,
	year = {2024},
	pages = {7436},
}

@article{calvanese_strinati_coherent_2020,
	title = {Coherent dynamics in frustrated coupled parametric oscillators},
	volume = {22},
	issn = {1367-2630},
	url = {https://dx.doi.org/10.1088/1367-2630/aba573},
	doi = {10.1088/1367-2630/aba573},
	language = {en},
	number = {8},
	urldate = {2025-05-15},
	journal = {New Journal of Physics},
	author = {Calvanese Strinati, Marcello and Aharonovich, Igal and Ben-Ami, Shai and Dalla Torre, Emanuele G and Bello, Leon and Pe’er, Avi},
	month = aug,
	year = {2020},
	pages = {085005},
}

@article{zhao_space-time_2025,
	title = {Space-time crystals from particle-like topological solitons},
	volume = {24},
	copyright = {2025 The Author(s)},
	issn = {1476-4660},
	url = {https://www.nature.com/articles/s41563-025-02344-1},
	doi = {10.1038/s41563-025-02344-1},
	language = {en},
	number = {11},
	urldate = {2025-12-19},
	journal = {Nature Materials},
	publisher = {Nature Publishing Group},
	author = {Zhao, Hanqing and Smalyukh, Ivan I.},
	month = nov,
	year = {2025},
	pages = {1802--1811},
}

@article{xu_space-time_2018,
	title = {Space-{Time} {Crystal} and {Space}-{Time} {Group}},
	volume = {120},
	url = {https://link.aps.org/doi/10.1103/PhysRevLett.120.096401},
	doi = {10.1103/PhysRevLett.120.096401},
	number = {9},
	urldate = {2025-12-19},
	journal = {Physical Review Letters},
	publisher = {American Physical Society},
	author = {Xu, Shenglong and Wu, Congjun},
	month = feb,
	year = {2018},
	pages = {096401},
}

@article{smits_observation_2018,
	title = {Observation of a {Space}-{Time} {Crystal} in a {Superfluid} {Quantum} {Gas}},
	volume = {121},
	url = {https://link.aps.org/doi/10.1103/PhysRevLett.121.185301},
	doi = {10.1103/PhysRevLett.121.185301},
	number = {18},
	urldate = {2025-12-19},
	journal = {Physical Review Letters},
	publisher = {American Physical Society},
	author = {Smits, J. and Liao, L. and Stoof, H. T. C. and van der Straten, P.},
	month = oct,
	year = {2018},
	pages = {185301},
}

@article{pauling_structure_1935,
	title = {The {Structure} and {Entropy} of {Ice} and of {Other} {Crystals} with {Some} {Randomness} of {Atomic} {Arrangement}},
	volume = {57},
	issn = {0002-7863},
	url = {https://doi.org/10.1021/ja01315a102},
	doi = {10.1021/ja01315a102},
	number = {12},
	urldate = {2025-12-22},
	journal = {Journal of the American Chemical Society},
	publisher = {American Chemical Society},
	author = {Pauling, Linus},
	month = dec,
	year = {1935},
	pages = {2680--2684},
}

@article{wannier_antiferromagnetism_1950,
	title = {Antiferromagnetism. {The} {Triangular} {Ising} {Net}},
	volume = {79},
	url = {https://link.aps.org/doi/10.1103/PhysRev.79.357},
	doi = {10.1103/PhysRev.79.357},
	number = {2},
	urldate = {2025-12-22},
	journal = {Physical Review},
	publisher = {American Physical Society},
	author = {Wannier, G. H.},
	month = jul,
	year = {1950},
	pages = {357--364},
}

@article{kang_complex_2014,
	title = {Complex {Ordered} {Patterns} in {Mechanical} {Instability} {Induced} {Geometrically} {Frustrated} {Triangular} {Cellular} {Structures}},
	volume = {112},
	url = {https://link.aps.org/doi/10.1103/PhysRevLett.112.098701},
	doi = {10.1103/PhysRevLett.112.098701},
	number = {9},
	urldate = {2025-12-22},
	journal = {Physical Review Letters},
	publisher = {American Physical Society},
	author = {Kang, Sung Hoon and Shan, Sicong and Košmrlj, Andrej and Noorduin, Wim L. and Shian, Samuel and Weaver, James C. and Clarke, David R. and Bertoldi, Katia},
	month = mar,
	year = {2014},
	pages = {098701},
}

@article{jorge_active_2024,
	title = {Active hydraulics laws from frustration principles},
	volume = {20},
	copyright = {2024 The Author(s)},
	issn = {1745-2481},
	url = {https://www.nature.com/articles/s41567-023-02301-2},
	doi = {10.1038/s41567-023-02301-2},
	language = {en},
	number = {2},
	urldate = {2025-12-22},
	journal = {Nature Physics},
	publisher = {Nature Publishing Group},
	author = {Jorge, Camille and Chardac, Amélie and Poncet, Alexis and Bartolo, Denis},
	month = feb,
	year = {2024},
	pages = {303--309},
}

@article{guo_non-orientable_2023,
	title = {Non-orientable order and non-commutative response in frustrated metamaterials},
	volume = {618},
	copyright = {2023 The Author(s), under exclusive licence to Springer Nature Limited},
	issn = {1476-4687},
	url = {https://www.nature.com/articles/s41586-023-06022-7},
	doi = {10.1038/s41586-023-06022-7},
	language = {en},
	number = {7965},
	urldate = {2025-12-23},
	journal = {Nature},
	publisher = {Nature Publishing Group},
	author = {Guo, Xiaofei and Guzmán, Marcelo and Carpentier, David and Bartolo, Denis and Coulais, Corentin},
	month = jun,
	year = {2023},
	pages = {506--512},
}

@article{wilczek_quantum_2012,
	title = {Quantum {Time} {Crystals}},
	volume = {109},
	url = {https://link.aps.org/doi/10.1103/PhysRevLett.109.160401},
	doi = {10.1103/PhysRevLett.109.160401},
	number = {16},
	urldate = {2025-12-23},
	journal = {Physical Review Letters},
	publisher = {American Physical Society},
	author = {Wilczek, Frank},
	month = oct,
	year = {2012},
	pages = {160401},
}

@article{bruno_comment_2013,
	title = {Comment on “{Quantum} {Time} {Crystals}”},
	volume = {110},
	url = {https://link.aps.org/doi/10.1103/PhysRevLett.110.118901},
	doi = {10.1103/PhysRevLett.110.118901},
	number = {11},
	urldate = {2025-12-23},
	journal = {Physical Review Letters},
	publisher = {American Physical Society},
	author = {Bruno, Patrick},
	month = mar,
	year = {2013},
	pages = {118901},
}

@article{zhang_observation_2017,
	title = {Observation of a discrete time crystal},
	volume = {543},
	copyright = {2017 Macmillan Publishers Limited, part of Springer Nature. All rights reserved.},
	issn = {1476-4687},
	url = {https://www.nature.com/articles/nature21413},
	doi = {10.1038/nature21413},
	language = {en},
	number = {7644},
	urldate = {2025-12-23},
	journal = {Nature},
	publisher = {Nature Publishing Group},
	author = {Zhang, J. and Hess, P. W. and Kyprianidis, A. and Becker, P. and Lee, A. and Smith, J. and Pagano, G. and Potirniche, I.-D. and Potter, A. C. and Vishwanath, A. and Yao, N. Y. and Monroe, C.},
	month = mar,
	year = {2017},
	pages = {217--220},
}

@article{kruss_nondispersive_2022,
	title = {Nondispersive {One}-{Way} {Signal} {Amplification} in {Sonic} {Metamaterials}},
	volume = {17},
	url = {https://link.aps.org/doi/10.1103/PhysRevApplied.17.024020},
	doi = {10.1103/PhysRevApplied.17.024020},
	number = {2},
	urldate = {2026-01-13},
	journal = {Physical Review Applied},
	publisher = {American Physical Society},
	author = {Kruss, Noah and Paulose, Jayson},
	month = feb,
	year = {2022},
	pages = {024020},
}

@article{melkani_space-time_2024,
	title = {Space-time symmetry and nonreciprocal parametric resonance in mechanical systems},
	volume = {110},
	url = {https://link.aps.org/doi/10.1103/PhysRevE.110.015003},
	doi = {10.1103/PhysRevE.110.015003},
	number = {1},
	urldate = {2026-01-13},
	journal = {Physical Review E},
	publisher = {American Physical Society},
	author = {Melkani, Abhijeet and Paulose, Jayson},
	month = jul,
	year = {2024},
	pages = {015003},
}

@article{veenstra_nonreciprocal_2025,
	title = {Nonreciprocal {Breathing} {Solitons}},
	volume = {15},
	url = {https://link.aps.org/doi/10.1103/nrv2-9h8z},
	doi = {10.1103/nrv2-9h8z},
	number = {3},
	urldate = {2026-01-16},
	journal = {Physical Review X},
	publisher = {American Physical Society},
	author = {Veenstra, Jonas and Gamayun, Oleksandr and Brandenbourger, Martin and van Gorp, Freek and Terwisscha-Dekker, Hans and Caux, Jean-Sébastien and Coulais, Corentin},
	month = aug,
	year = {2025},
	pages = {031045},
}

@article{kongkhambut_observation_2022,
	title = {Observation of a continuous time crystal},
	volume = {377},
	url = {https://www.science.org/doi/10.1126/science.abo3382},
	doi = {10.1126/science.abo3382},
	number = {6606},
	urldate = {2026-01-16},
	journal = {Science},
	publisher = {American Association for the Advancement of Science},
	author = {Kongkhambut, Phatthamon and Skulte, Jim and Mathey, Ludwig and Cosme, Jayson G. and Hemmerich, Andreas and Keßler, Hans},
	month = aug,
	year = {2022},
	pages = {670--673},
}

@article{bello_persistent_2019,
	title = {Persistent {Coherent} {Beating} in {Coupled} {Parametric} {Oscillators}},
	volume = {123},
	url = {https://link.aps.org/doi/10.1103/PhysRevLett.123.083901},
	doi = {10.1103/PhysRevLett.123.083901},
	number = {8},
	urldate = {2026-01-16},
	journal = {Physical Review Letters},
	publisher = {American Physical Society},
	author = {Bello, Leon and Calvanese Strinati, Marcello and Dalla Torre, Emanuele G. and Pe’er, Avi},
	month = aug,
	year = {2019},
	pages = {083901},
}

@book{dauxois_physics_2010,
	address = {Cambridge},
	title = {Physics of {Solitons}},
	isbn = {978-0-521-14360-8},
	language = {Anglais},
	publisher = {Cambridge University Press},
	author = {Dauxois, Thierry},
	year = {2010},
}

@article{frenkel_theory_1939,
	title = {On the theory of plastic deformation and twinning},
	volume = {1},
	url = {https://cds.cern.ch/record/431595},
	urldate = {2026-03-10},
	journal = {Izv. Akad. Nauk, Ser. Fiz.},
	author = {Frenkel, J and Kontorova, T},
	year = {1939},
	pages = {137--149},
}

@article{vekslerchik_discrete_1992,
	title = {Discrete nonlinear {Schrodinger} equation under nonvanishing boundary conditions},
	volume = {8},
	issn = {0266-5611},
	url = {https://doi.org/10.1088/0266-5611/8/6/007},
	doi = {10.1088/0266-5611/8/6/007},
	language = {en},
	number = {6},
	urldate = {2026-02-17},
	journal = {Inverse Problems},
	author = {Vekslerchik, V. E. and Konotop, V. V.},
	month = dec,
	year = {1992},
	pages = {889},
}

@book{ablowitz_discrete_2003,
	title = {Discrete and {Continuous} {Nonlinear} {Schrödinger} {Systems}},
	isbn = {978-0-511-54670-9},
	url = {http://dx.doi.org/10.1017/CBO9780511546709},
	doi = {10.1017/cbo9780511546709},
	publisher = {Cambridge University Press},
	author = {Ablowitz, M. J. and Prinari, B. and Trubatch, A. D.},
	month = dec,
	year = {2003},
}

@article{prinari_discrete_2016,
	title = {Discrete solitons of the focusing {Ablowitz}-{Ladik} equation with nonzero boundary conditions via inverse scattering},
	volume = {57},
	issn = {0022-2488},
	url = {https://doi.org/10.1063/1.4961160},
	doi = {10.1063/1.4961160},
	number = {8},
	urldate = {2026-03-10},
	journal = {Journal of Mathematical Physics},
	author = {Prinari, Barbara},
	month = aug,
	year = {2016},
	pages = {083510},
}

@article{cai_perturbation_1996,
	title = {Perturbation theories of a discrete, integrable nonlinear {Schr}{\textbackslash}textbackslashtextbackslashtextbackslash"odinger equation},
	volume = {53},
	url = {https://link.aps.org/doi/10.1103/PhysRevE.53.4131},
	doi = {10.1103/PhysRevE.53.4131},
	number = {4},
	urldate = {2026-02-25},
	journal = {Physical Review E},
	publisher = {American Physical Society},
	author = {Cai, David and Bishop, A. R. and Grønbech-Jensen, Niels},
	month = apr,
	year = {1996},
	pages = {4131--4136},
}

@article{mathew_synthetic_2020,
	title = {Synthetic gauge fields for phonon transport in a nano-optomechanical system},
	volume = {15},
	copyright = {2020 The Author(s), under exclusive licence to Springer Nature Limited},
	issn = {1748-3395},
	url = {https://www.nature.com/articles/s41565-019-0630-8},
	doi = {10.1038/s41565-019-0630-8},
	language = {en},
	number = {3},
	urldate = {2026-04-20},
	journal = {Nature Nanotechnology},
	publisher = {Nature Publishing Group},
	author = {Mathew, John P. and Pino, Javier del and Verhagen, Ewold},
	month = mar,
	year = {2020},
	pages = {198--202},
}

@article{miyake_realizing_2013,
	title = {Realizing the {Harper} {Hamiltonian} with {Laser}-{Assisted} {Tunneling} in {Optical} {Lattices}},
	volume = {111},
	url = {https://link.aps.org/doi/10.1103/PhysRevLett.111.185302},
	doi = {10.1103/PhysRevLett.111.185302},
	number = {18},
	urldate = {2026-04-20},
	journal = {Physical Review Letters},
	publisher = {American Physical Society},
	author = {Miyake, Hirokazu and Siviloglou, Georgios A. and Kennedy, Colin J. and Burton, William Cody and Ketterle, Wolfgang},
	month = oct,
	year = {2013},
	pages = {185302},
}

@article{mancini_observation_2015,
	title = {Observation of chiral edge states with neutral fermions in synthetic {Hall} ribbons},
	volume = {349},
	url = {https://www.science.org/doi/10.1126/science.aaa8736},
	doi = {10.1126/science.aaa8736},
	number = {6255},
	urldate = {2026-04-20},
	journal = {Science},
	publisher = {American Association for the Advancement of Science},
	author = {Mancini, M. and Pagano, G. and Cappellini, G. and Livi, L. and Rider, M. and Catani, J. and Sias, C. and Zoller, P. and Inguscio, M. and Dalmonte, M. and Fallani, L.},
	month = sep,
	year = {2015},
	pages = {1510--1513},
}

@article{veenstra_adaptive_2025,
	title = {Adaptive locomotion of active solids},
	volume = {639},
	copyright = {2025 The Author(s), under exclusive licence to Springer Nature Limited},
	issn = {1476-4687},
	url = {https://www.nature.com/articles/s41586-025-08646-3},
	doi = {10.1038/s41586-025-08646-3},
	language = {en},
	number = {8056},
	urldate = {2026-04-20},
	journal = {Nature},
	publisher = {Nature Publishing Group},
	author = {Veenstra, Jonas and Scheibner, Colin and Brandenbourger, Martin and Binysh, Jack and Souslov, Anton and Vitelli, Vincenzo and Coulais, Corentin},
	month = mar,
	year = {2025},
	pages = {935--941},
}

@article{slim_programmable_2025,
	title = {Programmable synthetic magnetism and chiral edge states in nano-optomechanical quantum {Hall} networks},
	volume = {16},
	copyright = {2025 The Author(s)},
	issn = {2041-1723},
	url = {https://www.nature.com/articles/s41467-025-62541-z},
	doi = {10.1038/s41467-025-62541-z},
	language = {en},
	number = {1},
	urldate = {2026-04-20},
	journal = {Nature Communications},
	publisher = {Nature Publishing Group},
	author = {Slim, Jesse J. and del Pino, Javier and Verhagen, Ewold},
	month = aug,
	year = {2025},
	pages = {7471},
}

@article{kivshar_dark_1998,
	title = {Dark optical solitons: physics and applications},
	volume = {298},
	issn = {0370-1573},
	shorttitle = {Dark optical solitons},
	url = {https://www.sciencedirect.com/science/article/pii/S0370157397000732},
	doi = {10.1016/S0370-1573(97)00073-2},
	number = {2},
	urldate = {2026-04-20},
	journal = {Physics Reports},
	author = {Kivshar, Yuri S. and Luther-Davies, Barry},
	month = may,
	year = {1998},
	pages = {81--197},
}

@article{belyansky_phase_2025,
	title = {Phase {Transitions} in {Nonreciprocal} {Driven}-{Dissipative} {Condensates}},
	volume = {135},
	url = {https://link.aps.org/doi/10.1103/gphr-d1bc},
	doi = {10.1103/gphr-d1bc},
	number = {12},
	urldate = {2026-05-28},
	journal = {Physical Review Letters},
	publisher = {American Physical Society},
	author = {Belyansky, Ron and Weis, Cheyne and Hanai, Ryo and Littlewood, Peter B. and Clerk, Aashish A.},
	month = sep,
	year = {2025},
	pages = {123401},
}

@article{apffel_frequency_2022,
	title = {Frequency {Conversion} {Cascade} by {Crossing} {Multiple} {Space} and {Time} {Interfaces}},
	volume = {128},
	url = {https://link.aps.org/doi/10.1103/PhysRevLett.128.064501},
	doi = {10.1103/PhysRevLett.128.064501},
	number = {6},
	urldate = {2026-05-28},
	journal = {Physical Review Letters},
	publisher = {American Physical Society},
	author = {Apffel, Benjamin and Fort, Emmanuel},
	month = feb,
	year = {2022},
	pages = {064501},
}

@article{wang_observation_2018,
	title = {Observation of {Nonreciprocal} {Wave} {Propagation} in a {Dynamic} {Phononic} {Lattice}},
	volume = {121},
	url = {https://link.aps.org/doi/10.1103/PhysRevLett.121.194301},
	doi = {10.1103/PhysRevLett.121.194301},
	number = {19},
	urldate = {2026-05-28},
	journal = {Physical Review Letters},
	publisher = {American Physical Society},
	author = {Wang, Yifan and Yousefzadeh, Behrooz and Chen, Hui and Nassar, Hussein and Huang, Guoliang and Daraio, Chiara},
	month = nov,
	year = {2018},
	pages = {194301},
}

@article{galiffi_broadband_2019,
	title = {Broadband {Nonreciprocal} {Amplification} in {Luminal} {Metamaterials}},
	volume = {123},
	url = {https://link.aps.org/doi/10.1103/PhysRevLett.123.206101},
	doi = {10.1103/PhysRevLett.123.206101},
	number = {20},
	urldate = {2026-05-28},
	journal = {Physical Review Letters},
	publisher = {American Physical Society},
	author = {Galiffi, E. and Huidobro, P. A. and Pendry, J. B.},
	month = nov,
	year = {2019},
	pages = {206101},
}

@article{fleury_floquet_2016,
	title = {Floquet topological insulators for sound},
	volume = {7},
	copyright = {2016 The Author(s)},
	issn = {2041-1723},
	url = {https://www.nature.com/articles/ncomms11744},
	doi = {10.1038/ncomms11744},
	language = {en},
	number = {1},
	urldate = {2026-05-28},
	journal = {Nature Communications},
	publisher = {Nature Publishing Group},
	author = {Fleury, Romain and Khanikaev, Alexander B. and Alù, Andrea},
	month = jun,
	year = {2016},
	pages = {11744},
}

@article{caloz_spacetime_2020,
	title = {Spacetime {Metamaterials}—{Part} {I}: {General} {Concepts}},
	volume = {68},
	issn = {1558-2221},
	shorttitle = {Spacetime {Metamaterials}—{Part} {I}},
	url = {https://ieeexplore.ieee.org/document/8858030},
	doi = {10.1109/TAP.2019.2944225},
	number = {3},
	urldate = {2026-05-28},
	journal = {IEEE Transactions on Antennas and Propagation},
	author = {Caloz, Christophe and Deck-Léger, Zoé-Lise},
	month = mar,
	year = {2020},
	pages = {1569--1582},
}

@article{delory_elastic_2024,
	title = {Elastic {Wave} {Packets} {Crossing} a {Space}-{Time} {Interface}},
	volume = {133},
	url = {https://link.aps.org/doi/10.1103/PhysRevLett.133.267201},
	doi = {10.1103/PhysRevLett.133.267201},
	number = {26},
	urldate = {2026-05-28},
	journal = {Physical Review Letters},
	publisher = {American Physical Society},
	author = {Delory, Alexandre and Prada, Claire and Lanoy, Maxime and Eddi, Antonin and Fink, Mathias and Lemoult, Fabrice},
	month = dec,
	year = {2024},
	pages = {267201},
}

@article{yves_symmetry-driven_2026,
	title = {Symmetry-driven artificial phononic media},
	volume = {11},
	copyright = {2025 Springer Nature Limited},
	issn = {2058-8437},
	url = {https://www.nature.com/articles/s41578-025-00860-9},
	doi = {10.1038/s41578-025-00860-9},
	language = {en},
	number = {2},
	urldate = {2026-05-28},
	journal = {Nature Reviews Materials},
	publisher = {Nature Publishing Group},
	author = {Yves, Simon and Fruchart, Michel and Fleury, Romain and Shmuel, Gal and Vitelli, Vincenzo and Haberman, Michael R. and Alù, Andrea},
	month = feb,
	year = {2026},
	pages = {156--180},
}

@misc{veenstra_wave_2025,
	title = {Wave coarsening drives time crystallization in active solids},
	url = {https://arxiv.org/abs/2508.20052v1},
	language = {en},
	urldate = {2026-05-28},
	journal = {arXiv.org},
	author = {Veenstra, Jonas and Binysh, Jack and Seinen, Vito and Naber, Rutger and Robledo-Poisson, Damien and Hunt, Andres and van Saarloos, Wim and Souslov, Anton and Coulais, Corentin},
	month = aug,
	year = {2025},
}

@book{strogatz_nonlinear_2015,
	address = {Boulder, CO},
	edition = {Second edition},
	title = {Nonlinear dynamics and chaos: with applications to physics, biology, chemistry, and engineering},
	isbn = {978-0-8133-4910-7},
	shorttitle = {Nonlinear dynamics and chaos},
	language = {en},
	publisher = {Westview Press, a member of the Perseus Books Group},
	author = {Strogatz, Steven H.},
	year = {2015},
}

@article{del_pino_non-hermitian_2022,
	title = {Non-{Hermitian} chiral phononics through optomechanically induced squeezing},
	volume = {606},
	copyright = {2022 The Author(s), under exclusive licence to Springer Nature Limited},
	issn = {1476-4687},
	url = {https://www.nature.com/articles/s41586-022-04609-0},
	doi = {10.1038/s41586-022-04609-0},
	language = {en},
	number = {7912},
	urldate = {2026-05-29},
	journal = {Nature},
	publisher = {Nature Publishing Group},
	author = {del Pino, Javier and Slim, Jesse J. and Verhagen, Ewold},
	month = jun,
	year = {2022},
	pages = {82--87},
}

@article{mahboob_electromechanical_2016,
	title = {An electromechanical {Ising} {Hamiltonian}},
	volume = {2},
	url = {https://www.science.org/doi/10.1126/sciadv.1600236},
	doi = {10.1126/sciadv.1600236},
	number = {6},
	urldate = {2026-05-29},
	journal = {Science Advances},
	publisher = {American Association for the Advancement of Science},
	author = {Mahboob, Imran and Okamoto, Hajime and Yamaguchi, Hiroshi},
	month = jun,
	year = {2016},
	pages = {e1600236},
}

@article{braginsky_parametric_2001,
	title = {Parametric oscillatory instability in {Fabry}–{Perot} interferometer},
	volume = {287},
	issn = {0375-9601},
	url = {https://www.sciencedirect.com/science/article/pii/S0375960101005102},
	doi = {10.1016/S0375-9601(01)00510-2},
	number = {5},
	urldate = {2026-05-29},
	journal = {Physics Letters A},
	author = {Braginsky, V. B. and Strigin, S. E. and Vyatchanin, S. P.},
	month = sep,
	year = {2001},
	pages = {331--338},
}

@article{de_bos_multimodal_2026,
	title = {Multimodal oscillator networks learn to solve a classification problem},
	volume = {2},
	copyright = {2025 The Author(s)},
	issn = {3059-3727},
	url = {https://www.nature.com/articles/s44455-025-00015-4},
	doi = {10.1038/s44455-025-00015-4},
	language = {en},
	number = {1},
	urldate = {2026-05-29},
	journal = {npj Metamaterials},
	publisher = {Nature Publishing Group},
	author = {de Bos, Daan and Serra-Garcia, Marc},
	month = jan,
	year = {2026},
	pages = {3},
}

@article{du_metamaterials_2026,
	title = {Metamaterials that learn to change shape},
	volume = {22},
	copyright = {2026 The Author(s), under exclusive licence to Springer Nature Limited},
	issn = {1745-2481},
	url = {https://www.nature.com/articles/s41567-026-03226-2},
	doi = {10.1038/s41567-026-03226-2},
	language = {en},
	number = {5},
	urldate = {2026-05-29},
	journal = {Nature Physics},
	publisher = {Nature Publishing Group},
	author = {Du, Yao and van Mastrigt, Ryan and Veenstra, Jonas and Coulais, Corentin},
	month = may,
	year = {2026},
	pages = {784--790},
}

@misc{lambert_nonlinear_2026,
	title = {Nonlinear stabilization of chiral modes in space-time modulated parametric oscillators},
	url = {http://arxiv.org/abs/2602.22513},
	doi = {10.48550/arXiv.2602.22513},
	urldate = {2026-06-10},
	publisher = {arXiv},
	author = {Lambert, Scott and Jaremko, Elise and Paulose, Jayson},
	month = may,
	year = {2026},
	note = {arXiv:2602.22513 [physics.class-ph]},
}

@article{mestre_network_2025,
	title = {Network of parametrically driven silicon nitride mechanical membranes},
	volume = {24},
	url = {https://link.aps.org/doi/10.1103/kx2w-czjf},
	doi = {10.1103/kx2w-czjf},
	number = {4},
	urldate = {2026-06-10},
	journal = {Physical Review Applied},
	publisher = {American Physical Society},
	author = {Mestre, Luis and Singh, Suyash and Margiani, Gabriel and Catalini, Letizia and Eichler, Alexander and Dumont, Vincent},
	month = oct,
	year = {2025},
	pages = {044072},
}

@article{nasari_observation_2026,
	title = {Observation of {Floquet} rotational super-radiance},
	volume = {655},
	copyright = {2026 The Author(s), under exclusive licence to Springer Nature Limited},
	issn = {1476-4687},
	url = {https://www.nature.com/articles/s41586-026-10725-y},
	doi = {10.1038/s41586-026-10725-y},
	language = {en},
	number = {8123},
	urldate = {2026-07-16},
	journal = {Nature},
	publisher = {Nature Publishing Group},
	author = {Nasari, Hadiseh and Moussa, Hady and Kasahara, Yoshiaki and Thielens, Arno and Alù, Andrea},
	month = jul,
	year = {2026},
	pages = {608--616},
}

@article{liu_controlled_2024,
	title = {Controlled pathways and sequential information processing in serially coupled mechanical hysterons},
	volume = {121},
	url = {https://www.pnas.org/doi/10.1073/pnas.2308414121},
	doi = {10.1073/pnas.2308414121},
	number = {22},
	urldate = {2026-07-21},
	journal = {Proceedings of the National Academy of Sciences},
	publisher = {Proceedings of the National Academy of Sciences},
	author = {Liu, Jingran and Teunisse, Margot and Korovin, George and Vermaire, Ivo R. and Jin, Lishuai and Bense, Hadrien and van Hecke, Martin},
	month = may,
	year = {2024},
	pages = {e2308414121},
}

@article{kwakernaak_counting_2023,
	title = {Counting and {Sequential} {Information} {Processing} in {Mechanical} {Metamaterials}},
	volume = {130},
	url = {https://link.aps.org/doi/10.1103/PhysRevLett.130.268204},
	doi = {10.1103/PhysRevLett.130.268204},
	number = {26},
	urldate = {2026-07-21},
	journal = {Physical Review Letters},
	publisher = {American Physical Society},
	author = {Kwakernaak, Lennard J. and van Hecke, Martin},
	month = jun,
	year = {2023},
	pages = {268204},
}

\end{document}


\author{Rupesh Mahore$^{(1)}$,Oleksandr Gamayun$^{(2)}$, Guillaume Noetinger$^{(3)}$, \\ Romain Fleury$^{(3)}$, Corentin Coulais$^{(2)}$, Benjamin Apffel$^{(4)}$}
\date{(1) Institute of Physics, Universiteit van Amsterdam, Science Park 904, 1098 XH Amsterdam, The Netherlands, \\ 
(2) London Institute for Mathematical Sciences, Royal Institution, 21 Albemarle St., London W1S 4BS \\
(3) Laboratory of Wave Engineering, EPFL, 1015 Lausanne, Switzerland \\
(4) CNRS, ENS de Lyon, LPENSL, UMR5672, 69342, Lyon cedex 07, France}

\maketitle
\tableofcontents

\section{Derivation of the envelope equation}
We are interested in the behavior of coupled parametric oscillators with a forcing that depends on the site. The evolution equation for the oscillator number $i$ is given by 
\begin{equation}
    \partial_t^2 \psi_i + 2\Gamma \omega_0 \partial_t \psi_i + \omega_0^2(1 + p\epsilon \cos{(2\omega t+\phi_i)}) (\psi_i - \alpha \psi_i^3) = k (\psi_{i+1} + \psi_{i-1} - 2\psi_i)
    \label{eq:psiTot}
\end{equation}
One can think of this equation as a nonlinear Sine-Gordon equation with an on-site parametric forcing. We now assume that the coupling goes as $k = \epsilon \kappa \omega_0^2$, $\omega = \omega_0 + \Delta \omega$ with small detuning $\Delta \omega = \epsilon \sigma \omega_0$, small dissipation $\Gamma = \epsilon \gamma$ and and we rescale $\psi \rightarrow \psi / \sqrt{\epsilon}$. This gives
\begin{equation}
    \partial_t^2 \psi_i +2\gamma\epsilon \omega_0 \partial_t \psi_i + \omega_0^2(1 + p\epsilon \cos{(2\omega t+\phi_i)}) (\psi_i - \epsilon \alpha \psi_i^3) = \epsilon \kappa \omega_0^2 (\psi_{i+1} + \psi_{i-1} - 2\psi_i)
\end{equation}
We perform the usual change of variable for averaging method
\begin{equation}
    \psi_i = A_i(t) e^{i \omega_0 t} + A_i^*(t) e^{-i \omega_0 t} \qquad \dot \psi_i = i\omega_0 (A_i(t) e^{i \omega_0 t} - A_i^*(t) e^{-i \omega_0 t})
\end{equation}
We find the two equations to be
\begin{align}
      \dot A_i e^{i\omega_0t} + c.c. &= 0 \\
    [i \omega_0 \dot A_i - \omega_0^2  A_i ]e^{i\omega_0t} + c.c. &= - \omega_0^2(1+p\epsilon \cos{(2\omega t+\phi_i)} (\psi-\epsilon\alpha \psi^3) \\
    &- 2i\gamma \omega_0^2 \epsilon  A_i(t) e^{i \omega_0 t}  +\epsilon \kappa \omega_0^2 (A_{i+1} + A_{i-1} - 2A_i)e^{i \omega_0t} + c.c.
\end{align}
At the end, this can be recast as
\begin{equation*}
    \begin{bmatrix}
        \dot A_i \\ \dot A_i^*
    \end{bmatrix} =
        \frac{\epsilon \omega_0}{2}
        \begin{bmatrix}
        \omega_0 e^{-i\omega_0 t} & -i e^{-i\omega_0 t}  \\ 
         \omega_0 e^{i\omega_0 t} & i e^{i\omega_0 t}  \\
    \end{bmatrix}
    \begin{bmatrix}
        0
        \\ 
         \alpha  \psi^3 -p\cos{(2\omega t+\phi_i}) \psi - 2 i\gamma A_i e^{i \omega_0t} + \kappa (A_{i+1} + A_{i-1} - 2A_i)e^{i \omega_0t}  + c.c.
    \end{bmatrix}
\end{equation*}
so that the equation on the amplitude is
\begin{equation*}
    \dot A_i =  i\frac{\epsilon \omega_0}{4} [pe^{i (2\Delta \omega t + \phi_i)} A_i^* - 6 \alpha \lvert A_i \rvert ^2 A_i + 4 i\gamma A_i - 2\kappa ( A_{i+1}+A_{i-1} - 2 A_i )] +  \text{fast oscillating} 
\end{equation*}
When averaging over one fast oscillation, we remove the fast oscillating term as $\dot A \sim \epsilon$ and thus varies slowly compared to $\omega_0$. The last step consists in resetting time as $u = \epsilon \omega_0 t$ to obtain
\begin{equation}
   -i\dot A_i = \frac{p}{4} e^{i(2\sigma u+\phi_i)} A_i^* -  \frac{3 \alpha}{2}|A_i|^2 A_i + i\gamma A_i - \frac{\kappa}{2} ( A_{i+1}+A_{i-1} - 2 A_i )
   \label{eq:generalAmplitude}
\end{equation}
The problem only depends on adimensioned damping and coupling, the forcing has been absorbed by redefinition of time and nonlinear strength by the amplitude rescaling. 

In the main text, we use a slightly different version of this equation. First, we set zero detuning $\sigma=0$. Second, we define $A_i=B_ie^{i\phi_i/2}$. Doing so, we obtain an equation on $B_i$ (that are also called $A_i$ in the main text)
\begin{equation}
   -i\dot B_i = \frac{p}{4} B_i^* -  \frac{3 \alpha}{2}|B_i|^2 B_i + i\gamma B_i - \frac{\kappa}{2} ( B_{i+1}e^{i(\phi_{i+1}-\phi_i)/2}+B_{i-1} e^{i(\phi_{i-1}-\phi_i)/2}- 2 B_i )
\end{equation}
which is used in the main text, as $(\phi_{i+1}-\phi_i)=W2\pi/N$. In the main text, we set $p=1$ which is always possible up to redefining $\epsilon$ and the other adimensional constants. The use of one or the other equation depends on the focus we want to set: the $A_i$ 's version set the focus on the onsite phase, while the $B_i's$ displays more clearly nonreciprocal coupling due to onsite phase gradient.

As we are here, it is also noteworthy to compute the continuous limit equation in an normalized form. For this, we start from Eq. (\ref{eq:generalAmplitude}) and we set again $\sigma=0$. Second, we write $A_j=A(ja)$ where $a$ is the length between two oscillators and assume that $A$ varies slowly over length $a$. 
\begin{equation}
   -i\partial_t A + \frac{\kappa a^2}{2} \partial_{xx}A + \frac{3 \alpha}{2}|A|^2 i= \frac{p}{4} e^{i\phi_i} A^*  + i\gamma A 
\end{equation}
We then reset the time scale $t'=\kappa a^2 t/2$ such that $\partial_t=\kappa a^2/2 \times\partial_{t'}$. Doing so, we get
\begin{equation}
   -i\partial_{t'} A + \partial_{xx}A + \frac{3 \alpha}{\kappa a^2}|A|^2 A= \frac{p}{2\kappa a^2} e^{i\phi(x)} A^*  + i\frac{2\gamma}{\kappa a^2}A 
\end{equation}
We then set $A=\sqrt{\frac{3\alpha}{2\kappa a^2}}\psi^*$ such that the equation on $\psi$ reads
\begin{equation}
   i\partial_{t'} \psi + \partial_{xx}\psi + 2s_\alpha |\psi|^2 \psi= \frac{p}{2\kappa a^2} e^{-i\phi(x)} \psi^*  - \frac{2i\gamma}{\kappa a^2}\psi 
\end{equation}
with $s_\alpha = \text{sign}(\alpha) = \pm 1$. We get a normalized non-linear Schrödinger equation on the left. Setting $\psi = \theta e^{-i\phi(x)/2}$, it can also be recast as
\begin{equation}
    i\partial_t \theta + \partial_{xx}\theta+2s_\alpha|\theta|^2 \theta = p'\theta^*-i\gamma'\theta+i(\partial_x\phi)\partial_x\theta + \frac{1}{4}\left((\partial_x\phi)^2+2i\partial_{xx}\phi \right) \theta
\end{equation}
with $p' = p/2\kappa a^2$ and $\gamma'=2\gamma/\kappa a^2$. Last, we can assume that $\phi(x) = Vx$ such that
\begin{equation}
    i\partial_t \theta + \partial_{xx}\theta+2s_\alpha|\theta|^2 \theta = p'\theta^*+\left(\frac{V^2}{4}-i\gamma\right)\theta+iV\partial_x\theta
\end{equation}
\section{A single parametric oscillator}
\subsection{Linear growth}
We first propose some reminders about the impact of nonlinearities on parametric oscillation. For this we consider a single oscillator which envelope obeys
\begin{equation}
    -i \partial_t A = \frac{e^{i\phi}}{4}A^* + i \gamma A_j - |A_j|^2A_j
\end{equation}
For small values of $A \ll 1$, nonlinearities can be neglected. In the linear case, the damping can always be removed by performing the substitution $ A \rightarrow A e^{-\gamma t}$. We are thus left with  
\begin{equation}
    -i \partial_t A = \frac{e^{i\phi}}{4}A^* 
\end{equation}
 Deriving the evolution equation a second time, we obtain
\begin{equation}
    \partial_{tt}A - \frac{1}{16}A=0 \rightarrow A(t) = A_0 e^{t/4}+B_0e^{-t/4}
\end{equation}
For large enough time $t$, the second term can be neglected. We are left with $A(t) = A_0e^{t/4}$ which can be plunged in the envelope equation to obtain
\begin{equation}
    -iA_0=e^{i\phi}A_0^*
\end{equation}
From this, we finally deduce that $A_0 = a_0e^{i(\phi/2+\pi/4)}$ with $a_0 \in \mathbb R$. Thus a single parametric oscillator gives solution of the form (where time has been redimensioned and damping added)
\begin{equation}
    \psi(t)=a_0e^{\epsilon\omega_0t/4} e^{-\Gamma \omega_0t}e^{i(\omega_0 t + \phi/2 + \pi/4)}
\end{equation}
There is also an exponentially decreasing solution $\sim e^{-\epsilon\omega_0 t/4}$ but it will be discarded as it is exponentially killed in time. We see that the oscillator is unstable for $\epsilon > 4\Gamma$ i.e. forcing must compensate for damping. For positive (resp. negative) values of $a_0$, the oscillator oscillates with phase $\phi/2+\pi/4$ (resp. $\phi/2+\pi/4+\pi$). Those are the two phase states discussed in the introduction.

\subsection{Nonlinear saturation}
As the amplitude exponentially grows, the nonlinearities will eventually become non negligible and finally lead to stationary amplitude $\partial_t A=0$. It can be computed as solution of
\begin{equation}
    \frac{e^{i\phi}}{4}A^* + i \gamma A = \frac{3\alpha}{2}|A_j|^2A_j
\end{equation}
For simplicity, we will assume that $\gamma$ is small such that we neglect it for now. Taking the modulus of the previous equation gives $|A_j|=1/\sqrt{6 |\alpha|} = a_L$ while looking at the phase gives $\arg A = \phi/2 [\pi]$. Therefore in the nonlinear regime, the steady solution reads
\begin{equation}
    \psi(t)=\pm a_Le^{i(\omega_0 t + \phi/2)}
\end{equation}
Depending on the sign in front, the oscillator oscillates with phase $\phi/2$ or $\phi/2 + \pi$. Compared to the linear case, we see that there is shift in the phase by $\pi/4$. Note that taking into account changes a bit this result.

\subsection{Energy balance}
We now propose an energy analysis of the nonlinear saturation, that will be used later. We can look at the energy variation that reads
\begin{equation}
    \partial_t |A|^2 = \frac{1}{2}\Re(ie^{i\phi}A^{*2}) 
\end{equation}
We see that the change of energy depends on the phase of $A^*$: $\Re(ie^{i\phi}A^{*2})$ is maximal for $\arg A = \phi/2 + \pi/4 [\pi]$ while it is zero for $\arg A = \phi/2 [\pi]$. We immediately recover the expected phases in the linear and nonlinear case. This gives an intuitive picture on the role of nonlinearities: as they grow, the phase of the complex amplitude varies, which slows down the energy injection until the later reaches 0. 

\section{Linear theory of the 1D chain}
We will first consider the simple case of no detuning ($\sigma = 0)$, no damping ($\gamma=0$) and no non-linearity ($\alpha = 0$). The evolution equation reads
\begin{equation}
   -i\dot A_i = \frac{1}{4} e^{i\phi_i} A_i^* +i\gamma A_i - \frac{\kappa}{2} ( A_{i+1}+A_{i-1} - 2 A_i )
\end{equation}
In order to simplify what follows, ae perform the change of variable $A_i = e^{-\gamma u} A_i$, which will remove the damping term while leaving the rest of the equation invariant. We thus assume $\gamma = 0$ up to adding this exponential decay in the solutions.

\subsection{Case N=2 oscillators}

We consider the case of 2 oscillators, which is already instructive. We get two evolution equations 
\begin{align}
   -i\dot A_1 &= \frac{1}{4} e^{i\phi_1} A_1^* - \kappa ( A_1 - A_2) \\
    -i\dot A_2 &= \frac{1}{4} e^{i\phi_2} A_2^* - \kappa ( A_2 - A_1)
\end{align}
We define $S = A_1 + A_2$ and $D=A_1 - A_2$ and assume $\phi_1 = 0$, $\phi_2=\pi$. The two evolution equations are
\begin{align}
   \ddot S - 2 i \kappa S - \frac{1}{16}S &= 0 \\
   \ddot D + 2 i \kappa D - \frac{1}{16}D &= 0 
\end{align}
which for $t \rightarrow \infty$ admits exponentially growing solutions of the form
\begin{equation}
    S(t) = S_0e^{(i \kappa+ \sigma) \epsilon \omega_0 t} \qquad D(t)=D_0 e^{(-i \kappa+ \sigma) \epsilon \omega_0 t} 
\end{equation}
with $\sigma = \sqrt{1-16\kappa^2}/4$ and $S_0,D_0$ a complex that depends on initial condition. There exists a relationship between those two constants. Indeed, one has
\begin{equation}
    -i \dot S = D^*/4
\end{equation}
such that plunging the previous solutions gives $S_0=\frac{D_0^*}{4(\kappa - i\sigma)}$. The next step is to remark that $|4(\kappa + i \sigma)| = 1$. By defining $\psi = - \text{arg}(\kappa + i\sigma)= -\arctan(\sigma/\kappa)$, and $D_0 = |D_0| e^{i(\psi_0-\psi/2)}$, we finally get
\begin{equation}
    S(t) = |D_0| e^{-i\psi_0} e^{(i \kappa+ \sigma) \epsilon \omega_0 t} \qquad D(t)=|D_0| e^{i\psi_0}e^{(-i \kappa+ \sigma) \epsilon \omega_0 t} 
\end{equation}
and we can recover the amplitude of each oscillator that goes as
\begin{equation}
   A_1(t) = A_0  \cos{(\kappa \epsilon\omega_0t-\psi_0)}e^{(\sigma-\gamma) \epsilon \omega_0 t} \qquad A_2(t) = i A_0  \sin{(\kappa \epsilon\omega_0t-\psi_0)}e^{(\sigma-\gamma) \epsilon \omega_0 t} 
\end{equation}
where we added again the damping term. There is a lot to learn from this expression
\begin{itemize}
    \item the enveloppe amplitude gets modulated due to coupling. In particular, it cancels every $T= \pi/ \kappa \epsilon \omega_0$ (the consinus cancels twice per period). 
    \item the canceling is associated with a sign change in the amplitude, which corresponds to a phase flip of the oscillator
    \item the phase flip are not synchronized but occur one after another with delay $T/4$
    \item the two oscillators are in phase quadrature due to the $i$ prefactor. This is due to phase locking of each oscillator with its forcing
\end{itemize}
There is also something remarkable: when one coupled the two parametric oscillators that were unstable, they can restabilize providing the coupling is strong enough ($\kappa > 1/4$). How is that possible since there is no dissipation in the system ? Actually, the energy injected by one oscillator  is in the exponentially killed Floquet mode of the other ! There is therefore an effective dissipation that occurs, but the energy is actually transmitted from one oscillator to the pump of the other !

On top of that, we see that the phase $\psi_0$ is unconstrained. It is in strong constrays with the parametric oscillation, which is constrained by the forcing. Hence we see that looking only at the slow amplitude, we retrieve some continuous symetry $t \rightarrow t + dt$.
\subsection{Case arbitrary N oscillators}
We set back the evolution equation in the form
\begin{equation}
   -i \partial_u \vec A = (P/2)^2 \vec A ^* + K \vec A
\end{equation}
where $P = \text{diag}(e^{i\phi_1/2}, ..., e^{i\phi_N/2}) $ and $K$ is the coupling matrix with non zero terms on the tri-diagonal $-\frac{\kappa}{2}(1, -2, 1)$. Note that $P^{-1} = P^\dagger  = P^*$, which will be used later. We slightly change this evolution equation by setting $\vec B = P^{-1} \vec A$ so that
\begin{equation}
   -i \partial_u \vec B = \vec B ^*/4 + Q \vec B
   \label{eq:B}
\end{equation}
with $Q = P^{-1} K P = Q^\dagger $. Taking the derivative of this expression, we end up with the rather simple expression
\begin{equation}
    \partial_{uu} B + i(Q^*-Q)\partial_u B + (Q^*Q-1/16) B = 0
\end{equation}
and we can write an explicit expression for $Q$ as a circular matrix
\begin{equation}
    Q=\frac{-\kappa}{2} \begin{bmatrix}
    -2 & e^{-i\Delta\phi_{12}/2} &0 &0&... &0 &e^{-i\Delta \phi_{1N}/2} \\
    e^{i\Delta \phi_{12}/2} & -2 & e^{-i\Delta \phi_{23}/2} & 0 & ... &0 & 0 \\
        0 &e^{i\Delta \phi_{23}/2} & -2 & e^{-i\Delta \phi_{34}/2} & 0 & ... &0 & \\
        ... & ...& ... & ...&  ... & ...&  ... & \\
        e^{i\Delta \phi_{1N}/2} & 0 &  ... & ...&0 & e^{i\Delta\phi_{N-1,N}/2} & -2
\end{bmatrix}
\end{equation}
where $\Delta \phi_{ij} = \phi_i - \phi_j$. Let us now assume that all the phase difference are the same, corresponding to $\phi_j= 2\pi (j-1) /n$ and $\Delta \phi_{i+1,i} = 2 \pi / n = \phi$. The eigenvalues of $Q$ are the eigenvalues of the coupling matrix $K$, which is the discrete Laplacian.  The latter admits the orthogonal eigenvectors
\begin{equation}
    v_k = \begin{bmatrix} 1 & u^k & u^{2k} ... & u^{(n-1)k} \end{bmatrix}^T = P^{2k} \begin{bmatrix} 1 & 1 & 1 ... & 1 \end{bmatrix}^T
\end{equation}
with $u = e^{2i\pi/n}$ is the primitive square root of unity. The associated eigenvalue is 
\begin{equation}
    \lambda_k = -\kappa(\cos\left( \frac{2k\pi}{N}\right)-1) = 2 \kappa \sin \left( \frac{k\pi}{N}\right)^2   
\end{equation}
The eigenvalues of $Q$ are the same and the corresponding eigenvectors are $\psi_k = P^{-1}v_k$. The eigenvector $\psi_k$ is also an eigenvector of $Q^*$ but the associated eigenvalue is $\lambda_{k-1}$. Indeed, one has $u = e^{i\phi}$ and thus
\begin{equation}
    P^{-2} v_k = P^{2(k-1)}\begin{bmatrix} 1 & 1 & 1 ... & 1 \end{bmatrix}^T = v_{k-1}
\end{equation}
such that
\begin{equation}
    Q^* \psi_k = P K P^{-2} v_k = P K v_{k-1} =\lambda_{k-1} P^{-1} P^2 v_{k-1} = \lambda_{k-1}P^{-1}v_k =  \lambda_{k-1}\psi_k
\end{equation}

The solutions of the differential equations are therefore of the form
\begin{equation}
    \vec B(t) = \sum_{k=1}^N C_k e^{X_k t}  \psi_k 
\end{equation}
where $X_k$ is the solution of the polynomial equation
\begin{equation}
    X_k^2 + i [\lambda_{k-1}- \lambda_{k}] X_k +  \lambda_{k-1}\lambda_{k} - 1/16 = 0
\end{equation}
which solutions are 
\begin{equation}
    X_k^\pm = -\frac{i}{2} [\lambda_{k-1}- \lambda_{k}] \pm \sqrt{\frac{1}{16} - \frac{(\lambda_{k-1}+ \lambda_{k})^2}{4} }
\end{equation}
To make it better, one can compute a bit further
\begin{align}
    X_k^\pm &= i\kappa \sin\left(\frac{(2k-1)\pi}{N} \right) \sin\left(\frac{\pi}{N} \right) \pm \frac{1}{4}\sqrt{1- 16\kappa^2 \left(1-\cos\left(\frac{(2k-1)\pi}{N} \right) \cos\left(\frac{\pi}{N} \right)\right)^2} \\
   & =i \mu_k \pm \sigma_k
\end{align}
and we now remark the nice properties
\begin{equation}
    X_{N+1-k}^\pm = (X_k^{\pm})^* = - i \mu_k \pm \sigma_k \quad \text{and} \quad \psi_{N+1-k}= \psi_{k}^* \quad \text{for} \quad1 \leq k \leq N
\end{equation}
If one waits long enough, the dominant term will be the one such that the growth rate $\sigma_k$ is maximal, which occurs for $k=1$ and $k=N$. We thus end up with
\begin{equation}
    \vec B(t) = (C_1 e^{i\mu_1 t}  \psi_1 + C_{N} e^{-i\mu_{1} t}  \psi_{1}^*)e^{\sigma_1 t} 
\end{equation}
We plug this expression in Eq. (\ref{eq:B}) that we project on $\psi_1$ (to remove the $\psi_1^*$ contributions thanks to orthogonality) and we get
\begin{equation}
    C_1 = \frac{C_N^*}{4(\mu_1 - i \sigma_1 - \lambda_1)} 
\end{equation}
As for the case $N=2$, we have $|4(\mu_1 - i \sigma_1 - \lambda_1)|^2 = 1$, so that $C_N^* = e^{i\phi_0}C_1$. From this, the behavior at long time of the amplitude reads (in redimensioned time)
\begin{equation}
    B_j(t) = C_0 \cos{(\mu_1 \epsilon \omega_0t + j\pi/N-\theta_0)} e^{\epsilon\omega_0 t \sqrt{1- 16\kappa^2\sin(\pi/N)^4}/4}
\end{equation}
with $\theta_0$ an aribitrary parameter. The amplitude of the $j^{th}$ oscillator cancels at $T_k^j = \frac{\pi}{\mu_1 \epsilon \omega_0}(k + \frac{1  }{2}-\frac{j}{n} )$. In particular, the time between two cancellations of the same oscillator is 
\begin{equation}
    T^j_{k+1}-T_k^j = \pi / \mu_1 \epsilon \omega_0
\end{equation}

and the time between cancellation of two consecutive oscillators is
\begin{equation}
    T^{j+1}_k-T^j_k = \frac{-\pi }{N\mu_1 \epsilon \omega_0} 
\end{equation} 
Using $\mu_1 = \kappa\sin(\pi/N)^2$, $\kappa = k/(\epsilon \omega_0^2)$, we finally get the velocity of the dislocation in site/sec)
\begin{equation}
\boxed{
    v_0 = -\frac{k}{\omega_0}\frac{N}{\pi}\sin(\pi/N)^2 \approx \frac{-k\pi}{N\omega_0}}
    \label{eq:linearSpeed}
\end{equation}
We see that the previous quantity is negative for $k>0$, which means that the $j+1^{th}$ oscillator switches phase just before the $j^{th}$. This proves that the phase dislocation goes against the phase gradient for positive coupling. Taking $k<0$ or changing the winding direction reverses the rotation direction  The dislocation moreover slows down as the number of oscillators increases. This can be understood as follow: for a very large number of oscillators, the neighbors oscillate almost in phase with each other and we thus expect the dislocation to take more time to jump from one to the other.

As for the case $N=2$, we also see that $\theta_0$ can be arbitrary, such that we retrive a continuous symetry $t \rightarrow t + dt$.
\section{Non-linear analysis}

\subsection{The role of boundary conditions}
We start from the non-linear evolution equation (we once again neglect damping for simplicity)
\begin{equation}
    -i \partial_t A_j = \frac{e^{i\phi_j}}{4}A_j^* - \frac{3\alpha}{2}|A_j|^2A_j-\frac{\kappa}{2}(A_{j+1}+A_{j-1}-2A_j)
\end{equation}
We seek for steady state $\partial_t A_j = 0$. From experimental results, we seek for solutions of the form $A_j = a_j e^{i\phi_j/2}$ with $a_j \in \mathbb R$. We find that
\begin{equation}
    a_j = 6\alpha|a_j|^2a_j+2 \kappa (a_{j+1}e^{i \theta_{j+1}}+a_{j-1}e^{i \theta_{j}}-2a_j)
\end{equation}
with $\theta_{j+1}=(\phi_{i+1}-\phi_i)/2=\pi/N$. Let us assume that the number of oscillators is large: in this case, taking $a_j=\pm1/\sqrt{6|\alpha|}=\pm a_L$ is solution assuming $\cos(\pi/N) \approx 1$. However, there is an issue due to periodic boundary condition. One has $A_{N}=A_0$ due to periodicity, but our ansatz gives $A_{N}=e^{i\phi_N/2}a_L=-a_L=-A_0$ as $\phi_{N}/2=\pi$. We therefore see that our solution is not consistently defined over the whole chain, a typical signature of non-orientable manifold. If one removes the periodic boundary condition and take an infinite chain, the solution is consistent.

\subsection{Energy analysis}
\subsubsection{Fully saturated model}
We now aim to write an energy equation. In what follows, we take $\phi_j= W\times 2j\pi/N$ with $W=\pm1$ the winding direction of the phase, and assume $\kappa > 0$ (positive coupling). We moreover assume that $N \gg 1$ such that $\phi_{j+1}-\phi_j = 2\pi/N \ll1$. Using the previous equation, we get
\begin{equation}
    \partial_t |A_j|^2 = \frac{1}{2} \Re(i e^{i \phi_j}A_j^{*2 }) - \kappa\Re(i (A_{j+1}+A_{j-1})A_j^*)
\end{equation}
The first term is the on-site injection of energy, that can go to zero due to nonlinearities as discussed above. The second term is a flux term toward neighbors. We will assume that all oscillators are close from saturation i.e. that the onsite term is negligible. We will refer to this approximation as "fully saturated model", for which we seek solutions of the form
\begin{equation}
    A_j(t) = a_j(t)e^{i \phi_j/2}
\end{equation}
where $a_j(t) \sim a_L$ is a real function of the order of the maximal amplitude $a_L$. Despite this is not well defined globally, due to PBC, it is well defined locally (i.e. for a small portion of the chain). Replacing the ansatz in the equation gives 
\begin{equation}
    \partial_t |a_j|^2 =  \kappa  a_j (a_{j+1}-a_{j-1}) \sin(W\pi/N)
    \label{eq:transferEnegySaturated}
\end{equation}
with $\Delta \phi_{j+1}=\phi_{j+1}-\phi_j$.  

Let us first assume that the steady state is locally reached i.e. $a_j = a_{j+1}=a_{j-1}=a_L$: all oscillators are in phase with saturated amplitude. In this case, we immediately see that $\partial_t|a_j|^2 = 0$, which is consistent: there is no energy flux toward neighbors in the steady state.

Let us now assume that the frustration is located between the site $j$ and $j+1$. We will moreover assume that all the oscillators are approximately saturated i.e. $a_{j-1}\sim a_j\sim a_L$ and $a_{j+1} \sim a_{j+2}\sim-a_L$, ensuring a phase difference of $\pi$ between oscillator $j$ and $j+1$. In this case, we get
\begin{align}
    \partial_t |a_j|^2 &\sim -2\kappa a_L^2\sin(W\pi/N) \\
    \partial_t |a_{j+1}|^2 &\sim 2\kappa a_L^2\sin(W\pi/N)
\end{align}
We therefore see that the presence of the phase defects leads to a flux of energy from one oscillator to the other. For $W>0$ (resp. $W<0$), the energy goes from $j$ to $j+1$ (resp. $j+1$ to $j$ ). The energy thus flows along the phase gradient, with a time scale (with redimensioned time to express it in seconds)
\begin{equation}
\boxed{
    \frac{1}{\tau} = 2\kappa \epsilon \omega_0\sin(W\pi/N) = 2 \frac{k}{\omega_0}\sin(W\pi/N) \sim v_0}
\end{equation}
which is the result used in the main text. The velocity is moreover expected to be of $\sim 1/\tau$ site/s . We retrieve the same order of magnitude $k\pi/(N\omega_0)$ already found in the linear analysis (\ref{eq:linearSpeed}). The prefactor 2 can be simply explained, as $\tau$ is the typical decrease time of energy (i.e. $|\psi|^2$), which is exactly twice the decrease time of $\psi$ considered in the linear case. More precisely, if $|\psi|^2 \sim e^{-t/\tau}$, then $|\psi|\sim e^{-t/2\tau}$, hence the prefactor 2 between the two approaches.

Let us assume for instance that $W>0$. We thus expect $a_j=a_L$ to decrease while $a_{j-1},a_{j+1}$ should remain of the order of $a_L$. From eq. (\ref{eq:transferEnegySaturated}), if $a_j$ goes to zero and becomes negative, $\partial_t |a_j|^2$ turns positive again. The change of sign in $a_j$ corresponds to a phase switch of $\pi$, after which the oscillator amplitude grows again. As a result, it will eventually reach its new steady state $a_j = -a_L$. We now have $a_{j+1}=a_j=-a_L$ while $a_{j-1}=a_L$: the phase dislocation has jumped between site $j$ and $j-1$. Therefore, the phase dislocation goes against the phase gradient ($W>0$ here).

\subsubsection{Freezing of the dislocation}

\begin{figure}
    \centering
    \includegraphics[width=15cm]{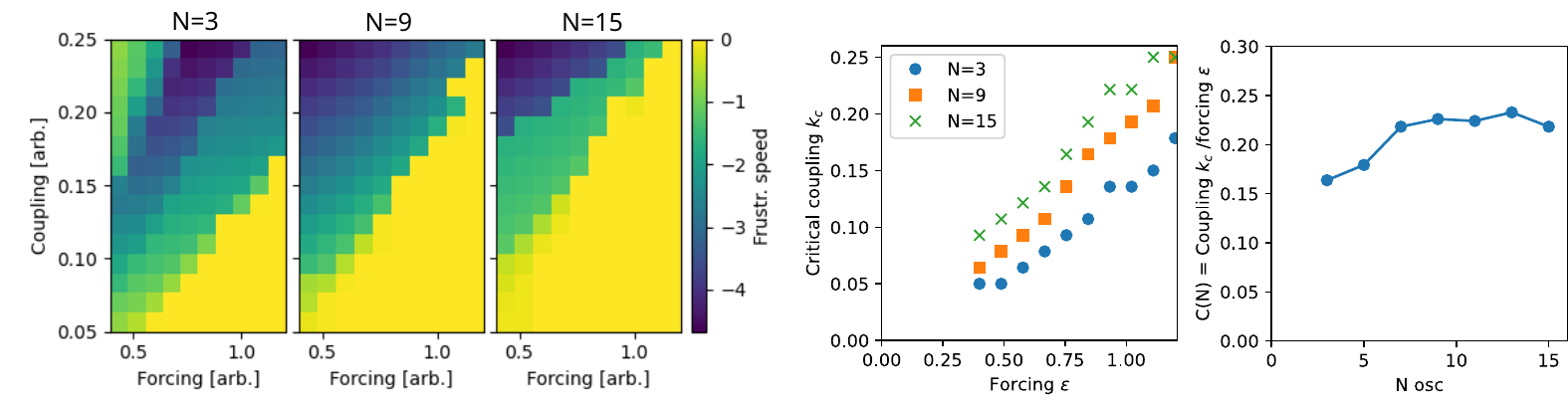}
    \caption{Numerical simulations of the nonlinear equation. (a) Velocity of the frustration for different values of forcing $\epsilon$ and coupling $k$. A clear transition between moving and nonmoving for $k=k_c(\epsilon)$ is visible. (b) Critical coupling $k_c(\epsilon)$ above which the frustration moves for different values of $N$. A linear fit allows to extract $k_c/\epsilon=C(N)$ (c) Critical ratio $C(N)$ as a function of $N$.}
    \label{fig:simuNL}
\end{figure}
In the previous analysis, we have used Eq. (\ref{eq:transferEnegySaturated}) but as $a_j$ decreases, we expect the phase of $A_j$ to vary (see section 2 for a single oscillator). This means our ansatz $A_j = a_j(t) e^{i \phi/2}$ will eventually break down for site $j$, and Eq. (\ref{eq:transferEnegySaturated}) will no longer be valid for $|a_j| \ll a_L$. We can nevertheless still assume that the neighboring oscillators are saturated: $A_{j+1} = - a_Le^{i\phi_{j+1}}$ and $A_{j-1} = a_Le^{i\phi_{j-1}}$. We also set $A_j(t) = B_j(t) e^{i\phi_j/2}$ where $B_j$ is now a complex variable. At small $t$, we have $B_j(t) \sim a_L$ so that it is a positive real number. Similarly, at long time, we expect $B_j \rightarrow \pm a_L$ depending whether a phase transition occured or not.

The energy balance now reads
\begin{align}
    \partial_t |B_j|^2 =& \frac{1}{2} \Re(i B_j^{*2 }) - 2\kappa a_L \sin\left(\frac{W\pi}{N}\right) \Re( B_j^*) \\
    =& \left[\Im(B_j) - 2\kappa a_L \sin\left(\frac{W\pi}{N}\right)\right] \Re( B_j)
\end{align}
Since $|\Im(B_j)|\sim a_L$ at most, we see that if $2\kappa \sin(W \pi/N) >1$ and $\Re(B_j)>0$, the energy will go down to zero, as the bracket will always be negative. The only way for the energy to grow again is to get $\Re(B_j)<0$, which indicates a phase switch. Remembering that $\kappa=k/(\epsilon\omega_0^2)$, this means that the dislocation must move if 
\begin{equation}
    k > \frac{\omega_0^2\epsilon}{2\sin(W\pi/N)} 
\end{equation}
The freezing occurs when the onsite pumping is strong enough to compensate the flux toward the neighbors generated by the frustration. It is this competition between onsite and coupling that give rise to this freezing/moving transition observed experimentally. The previous criteria gives the minimal ratio $k/\epsilon = C(N)$ below which the coupling is for sur stronger than the pump.

We have performed numerical simulations for different values of $N, \epsilon,k$ while $W=1$ while $\Gamma=0.05$ and $\omega_0=1$. In each case, we measured the velocity of the frustration. The results are presented in Fig. \ref{fig:simuNL}. For each length of the chain, we clearly see a transition between moving and nonmoving frustration. The transition occurs for $k_c(\epsilon) \approx C(N) \epsilon$, where $C(N)$ slightly increases before saturating around $0.22$ at large value of $N$. Our numerical simulations confirm that the scaling law found above $k_c = C(N) \epsilon$ is relevant. Nevertheless, the bound found above for $C(N)$ is thus not very precise since $1/\sin(W\pi/N) \rightarrow \infty$ as $N$ grows. Obtaining an actual estimate of the freezing/moving transition would require more precise results on the behavior of $B_j$, which is a difficult task.

In the next subsection, we propose an alaysis in terms of solitons. One obtains that the discretness of the lattice pins the kink for coupling such that
\begin{equation}
    k<k_c=\sqrt{(p\epsilon/4)^2-\Gamma^2}
\end{equation}
In our numerical simulations, one has $\epsilon>0.4 \gg\Gamma$. Therefore, the criteria becomes $k_c = \epsilon/4$ i.e. $C(N)=0.25$. It is of the good order of magnitude and in good agreement for larger values of $N$.

\section{Soliton analysis}
\subsection{Toward integrable equation}
We now propose an analysis of the dislocation in terms of solitons, to derive the result discussed above. For completeness, we shall start again from the full Eq. (1) and perform a slightly different version of envelope analysis. We assume that
\begin{itemize}
    \item perform rescaling $\psi\to \psi/(\sqrt{\omega_0 \alpha})$
    \item ignore detuning $\omega \approx \omega_0$
    \item discard terms proportional to $O(\epsilon^2)$. 
\end{itemize}
After implementing these rescaling, the equation becomes
\begin{equation}
    \partial_t^2 \psi_n - \epsilon \kappa \omega_0^2 (\psi_{n+1} + \psi_{n-1} - 2\psi_n)
    + \omega_0^2 \bigl(1 + p\epsilon \cos(2\omega t + \phi_n)\bigr)\psi_n
    + \omega_0  \psi_n^3
    + 2\gamma \epsilon \omega_0 \partial_t \psi_n = 0.
\end{equation}
We then put
\begin{equation}
    \psi_n = A_n(t) e^{-i\omega_0 t-i\phi_n/2}+ 
    A^*_n(t) e^{i\omega_0 t+i\phi_n/2}
\end{equation}
Assuming that $\dot{A}_n(t) \ll \omega_0 A_n(t)$ and $\phi_n = 2\pi W n/N$ we obtain 
\begin{equation}\label{ant}
    2i \dot{A}_n(t) + \epsilon \kappa\omega_0\left(A_{n+1}e^{\pi i W/N}+A_{n-1}e^{-\pi i W/N}-2A_n\right)- \frac{\omega_0 p \epsilon}{2} A_n^*(t) -3A_n^2 A_n^* + 2 i \omega_0 \gamma \epsilon A_n = 0 
\end{equation}
Let us perform a set of transformations to this equation. First we rescale $A_n$ by the constant complex number 
\begin{equation}
    A_n(t) = A e^{i\delta} \alpha_n(t), 
\end{equation}
which transforms Eq. \eqref{ant} into 
\begin{equation}
        2i \dot{\alpha}_n + \epsilon \kappa\omega_0\left(\alpha_{n+1}e^{\pi i W/N}+\alpha_{n-1}e^{-\pi i W/N}-2\alpha_n\right)- \frac{\omega_0 p \epsilon}{2} e^{-2i\delta} \alpha_n^* -3A^2\alpha_n^2 \alpha_n^* + 2 i \omega_0 \gamma \epsilon \alpha_n = 0 
\end{equation}
If we choose the phase $\delta$ such that 
\begin{equation}
    e^{-2i\delta} = -\sqrt{1 - \left( \frac{4\gamma}{p}\right)^2} + \frac{4i\gamma}{p}
\end{equation}
we obtain identically,
\begin{multline}
            2i \dot{\alpha}_n + \epsilon \kappa\omega_0\cos\frac{\pi W}{N}\left(\alpha_{n+1}+\alpha_{n-1}-2\alpha_n\right) -3A^2\alpha_n^2 \alpha_n^*  +
            \epsilon\omega_0\left(2\kappa \left(\cos\frac{\pi W}{N}-1\right)+\frac{p}{2}\sqrt{1 - \left( \frac{4\gamma}{p}\right)^2} \right) \alpha_n  \\ +
            \left(\frac{\omega_0 p \epsilon}{2} \sqrt{1 - \left( \frac{4\gamma}{p}\right)^2} -2 i \omega_0 \gamma \epsilon \right)(\alpha_n^*-\alpha_n) + i \epsilon\kappa\omega_0 \sin \frac{\pi W}{N} (\alpha_{n+1}-\alpha_{n-1}) = 0
\end{multline}
 Further, we can use rescaling of the time to fix the coefficients between first two terms and pick the parameter $A$ to fix the coefficient at the non-linear term. Overall, we claim that if we perform transformation 
\begin{equation}
 A_n(t) = e^{i\delta} \sqrt{\frac{2\epsilon \kappa \omega_0}{3} \cos \frac{\pi W}{N} } u_n\left(\tau\right),\qquad \tau =   t \frac{\epsilon \kappa \omega_0}{2} \cos \frac{\pi W}{N},  
\end{equation}
then Eq. \eqref{ant} takes the form of the perturbed discrete non-linear Schroedinger equation (dNLS)
\begin{equation}\label{uu}
\boxed{
    i \dot{u}_n + u_{n+1}+u_{n-1}-2u_n + 2u_n (\rho^2 -|u_n|^2) = R[u]}
\end{equation}
with 
\begin{equation}\label{RU}
    R[u]= \Omega (u_n - u^*_n) - i (u_{n+1}-u_{n-1})\tan \frac{\pi W}{N} 
\end{equation}
and parameters given by 
\begin{equation}
    \rho^2 =1-\frac{1}{\cos \frac{\pi W}{N}}+ \frac{\sqrt{p^2-16\gamma^2}}{4\kappa \cos \frac{\pi W}{N}},\qquad
    \Omega = \frac{\sqrt{p^2-16\gamma^2}}{2\kappa \cos \frac{\pi W}{N}} - \frac{2i\gamma }{\kappa \cos \frac{\pi W}{N}}.
    \label{eq:defRho}
\end{equation}
The advantage of this presentation is that after removing the phase $\delta$ the profile will be mostly real during the time evolution.

To gain insight into the type of excitations induced by the nonlinearity, we consider the continuum limit of Eq.~\eqref{uu}. Namely, in the limit \( u_n(t) \to \psi(x,t) \), Eq.~\eqref{uu} reduces to a perturbed nonlinear Schr\"odinger equation (NLS),
\begin{equation}\label{defoc2}
i \partial_t \psi + \partial_x^2 \psi + 2\psi \bigl(\rho^2 - |\psi|^2\bigr) = R[\psi] = 2i \tilde{\Omega} \operatorname{Im}\psi + i \tilde{v} \partial_x \psi.
\end{equation}
Here, the tilde denotes possible renormalizations of parameters arising in the continuum limit. It is immediately apparent that the last term in the perturbation \( R[\psi] \) can be eliminated via the Galilean transformation \( x \to x + \tilde{v} t \). Furthermore, Eq.~\eqref{defoc2} admits the exact dark soliton (kink) solution
\begin{equation}\label{exact1}
\psi(x,t) = \pm \rho \tanh\bigl(\rho(x + \tilde{v} t)\bigr).
\end{equation}

Numerical simulations indicate that, starting from random initial conditions, solutions of Eq.~\eqref{uu} evolve toward configurations consisting of kinks similar to \eqref{exact1}. In fact, if we replace the left-hand side of Eq.~\eqref{defoc2} with its integrable discretization, known as the Ablowitz--Ladik (AL) system \cite{vekslerchik_discrete_1992,ablowitz_discrete_2003,prinari_discrete_2016},
\begin{equation}\label{al1}
i \dot{u}_n + u_{n+1} + u_{n-1} - 2u_n + 2u_n \rho^2 - |u_n|^2 (u_{n+1} + u_{n-1}) = R[u],
\end{equation}
then the analog of Eq.~\eqref{exact1} is the exact (unperturbed) solution
\begin{equation}\label{un}
    u_n(t) = \rho \tanh\bigl(B (n - x_0)\bigr), \qquad B = \frac{1}{2} \ln \frac{1+\rho}{1-\rho}.
\end{equation}

For low values of forcing, numerical simulations of Eq.~\eqref{al1} confirm that this system closely approximates the continuum solution \eqref{exact1}, even when the perturbation term \( R[u] \) in \eqref{RU} is included. On the other hand, DNS of Eq.~\eqref{uu} at higher forcing reveal qualitatively different behavior for certain parameter regimes, resulting in pinning of kinks even when \( W \neq 0 \). Such behavior is not observed for DNS of the continuous equation.

This demonstrates that the pinning is intrinsically tied to the discreteness of the system. Notably, the solution \eqref{un}, in contrast to the continuous case \eqref{exact1}, exists only for \( 0 < \rho < 1 \), in order to define $B$ in \eqref{un}. Consequently, as \( \rho \to 1 \), lattice effects become increasingly significant, and the discrepancy between the Ablowitz--Ladik model and the original system \eqref{uu} grows. In this regime, the kink solution is no longer existing, such that the phase dislocation remains pinned by the lattice potential.

In the limit \( N \gg 1 \), one can estimate the critical value for which $\rho \sim 1$, which yields 
\begin{equation}
    \boxed{p_c = 4 \sqrt{\kappa^2 + \gamma^2}.}
\end{equation}
This way, for $p>p_c$ we expect the pinning of the kinks.
Expressed in terms of dimensional variables, this is equivalent to
\begin{equation}
   \boxed{ k > k_c = \sqrt{\frac{(p\epsilon)^2}{16} - \Gamma^2}}
\end{equation}
which is in qualitative agreement with the numerical results shown in Fig.~\ref{fig:simuNL}.

\subsection{SNIC scenario due to discretness of lattice}
We further study the role of discreteness, we replace as before the left-hand side of Eq. \eqref{uu} with the integrable discrete NLS equation, namely the Ablowitz–Ladik (AL) system \cite{vekslerchik_discrete_1992,ablowitz_discrete_2003,prinari_discrete_2016},
\begin{equation}
i \dot{u}_n + u_{n+1}+u_{n-1}-2u_n + 2u_n \rho^2 -|u_n|^2(u_{n+1}+u_{n-1}) = R[u].
\end{equation}
We then define the sum of squared amplitudes for the perturbed discrete NLS model \eqref{uu},
\begin{equation}
    Q \equiv \sum\limits_{n} \bigl(\rho^2 - |u_n|^2\bigr).
\end{equation}
This quantity is also known as the power, or the number of particles \cite{kivshar_dark_1998}, and was already used above to perform energy balance. Its time evolution is given by
\begin{equation}
    \frac{dQ}{dt} = -\sum\limits_{n}\frac{d |u_n|^2}{dt} = \sum\limits_{n} \left(j_n - j_{n-1} + i\bigl(u_n^* R_n - u_n R_n^*\bigr)\right),
\end{equation}
where
\begin{equation}
    j_n = 2 \operatorname{Im}\bigl(u_n^* u_{n+1}\bigr).
\end{equation}
This way, the number of particles is conserved for the unperturbed model \( R_n = 0 \). In fact, the terms proportional to \( \tan(\pi W/N) \) in \eqref{RU} do not spoil conservation, and the full time evolution can be written as
\begin{equation}\label{25}
    \frac{dQ}{dt} = \sum\limits_{n} \left(J_n - J_{n-1} - 2 (u_n^* \Omega + u_n \Omega^*) \operatorname{Im} u_n \right),
\end{equation}
with
\begin{equation}
    J_n = 2 \operatorname{Im}\bigl(u_n^* u_{n+1}\bigr) + 2 \tan\frac{\pi W}{N} \operatorname{Re}\bigl(\rho^2 - u_n^* u_{n+1}\bigr).
\end{equation}
For purely real profiles, this implies
\begin{equation}
    \frac{dQ}{dt} = 0,
\end{equation}
even in the presence of perturbations.

\begin{figure}
    \centering
    \includegraphics[width=\linewidth]{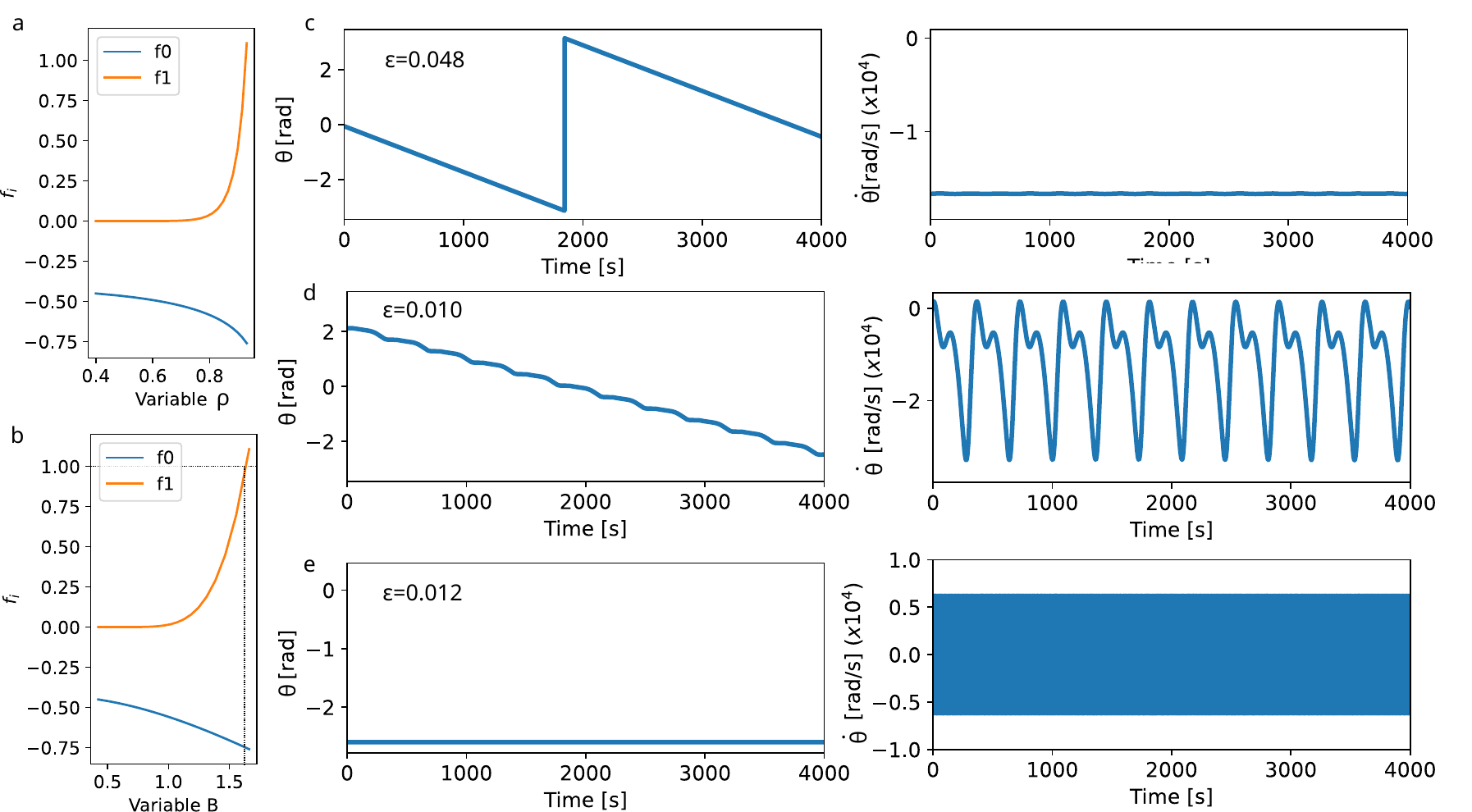}
    \caption{SNIC bifurcation. (a) Fonctions $f_0$ and $f_1$ as a function of $\rho$ and (b) as a function of $B$. (c) Angular position $x_0(t)$ and velocity $\dot x_0$ of the soliton on the circular chain, defined from \eqref{eq:defXsoliton} and computed from DNS with $\omega_0=1$, $\Gamma=0.01$, $k=0.02$. For $\epsilon=0.048$, propagation is regular. For $\epsilon=0.01$, each site acts as a "bump" and a step-like trajectory appears. It is confirmed by the velocity plot that exhibits periodic variation, in qualitative agreement with Eq. \ref{eq:SNICf}. However, note that $\dot \theta$ turns positive for some time, which is not predicted for a SNIC bifurcation. (e) For $\epsilon =0.012$, the dislocation is stuck.}
    \label{fig:S3}
\end{figure}

Numerical simulations suggest that the solution can be approximated by an adiabatically deformed version of the exact profile \eqref{un}, namely
\begin{equation}\label{prof}
    u_n(t) = \rho \tanh\bigl(B (n - x_0)\bigr),
\end{equation}
where the parameters vary slowly in time, \( B = B(t) \) and \( x_0 = x_0(t) \). The total power \( Q \) can then be computed explicitly,
\begin{equation}
    Q = \frac{2\rho^2}{B} + \frac{4\pi^2 \rho^2}{B^2} \sum\limits_{k=1}^\infty \frac{k \cos(2\pi k x_0)}{\sinh(\pi^2 k/B)}
    \approx \frac{2\rho^2}{B} + \frac{8\pi^2 \rho^2}{B^2} \cos(2\pi x_0)\, e^{-\pi^2/B}.
\end{equation}
Conservation of \( Q \) implies \( \dot{B} = \mathcal{O}(e^{-\pi^2/B}) \). On the other hand, the soliton velocity satisfies \( \dot{x}_0 = \mathcal{O}(1) \). To estimate it, we consider the center of mass of the squared amplitudes \cite{cai_perturbation_1996},
\begin{equation}
    X \equiv \sum\limits_{n} n \bigl(\rho^2 - |u_n|^2\bigr).
    \label{eq:defXsoliton}
\end{equation}
On the one hand, we obtain
\begin{equation}
    X \approx x_0 Q + \frac{2\pi^2 \rho^2}{B} \sin(2\pi x_0)\, e^{-\pi^2/B} \left(1 - \frac{\pi^2}{B}\right),
\end{equation}
which yields
\begin{equation}
    \frac{dX}{dt} \approx \frac{dx_0}{dt} \left(Q + \frac{4\pi^3 \rho^2}{B} \cos(2\pi x_0)\, e^{-\pi^2/B} \left(1 - \frac{\pi^2}{B}\right)\right).
\end{equation}
On the other hand,
\begin{equation}
    \frac{dX}{dt} = \sum_n n (J_n - J_{n-1}) = -\sum_n J_n = -4 \rho^2 \tan\frac{\pi W}{N} \coth(B),
\end{equation}
which leads to
\begin{equation}\label{x0}
    \boxed{\frac{dx_0}{dt} \approx -2B \coth(B) \tan\frac{\pi W}{N}
    \left(
    1 - \frac{2\pi^2}{B} \cos(2\pi x_0)\, (2 + \pi - \pi^3 B)\, e^{-\pi^2/B}
    \right)}
\end{equation}
which is of the form
\begin{equation}
    \dot x_0 = f_0(p) + f_1(p) \cos{(2\pi x)}
    \label{eq:SNICf}
\end{equation}
The dependency of $f_0$ and $f_1$ with respect to the pump is encompassed in $B$, which depends on $\rho$ given by \eqref{eq:defRho}. In particular, $\rho$ (and thus $f_0,f_1$) only depends on the pump $p$ in the limit $W/N \ll 1$. It is plotted in Fig. \ref{fig:S3}a-b. In the main text, we perform the approximation $f_0(p) \approx f_0(0) = -2\tan(W\pi/N) $. As shown in Fig. \ref{fig:S3}a, it is reasonable except for $\rho$ very close from 1. We finally obtain the equation used in the main text
\begin{equation}
     \dot x_0 = -2 \tan\left( \frac{W\pi}{N} \right)[1 + f_1(p) \cos{(2\pi x)}]
\end{equation}
Remembering that time was adimensioned, the final result in actual units reads
\begin{equation}
    \frac{dx_0}{dt} = \frac{-k}{\omega_0}\sin\left(\frac{\pi W}{N} \right) [1 + f_1(p)\cos(2\pi x)]
\end{equation}

Close from the instability threshold, one has $\rho^2 \ll1$ such that $B \approx 0$. In this case, one can approximate the previous equation as
\begin{equation}
\boxed{
    \frac{dx_0}{dt} \approx - \frac{k}{\omega_0} \sin\left(\frac{\pi W}{N} \right), \qquad \rho \ll1 }
    \label{eq:velocityLowForcing}
\end{equation}
which is similar to the result obtained with our energy argument, as well as the result obtained from the linear analysis (\ref{eq:linearSpeed}). On the other hand, when $\rho \rightarrow1^-$, one has $B\rightarrow + \infty$ such that
\begin{equation}
\boxed{
    \frac{d x_0}{dt} \approx-2B \tan\left( \frac{W\pi}{N}\right)(1+2\pi^5\cos(2\pi x_0) ), \qquad \rho \sim1^- }
\end{equation}
and we have indeed obtained a SNIC transition since $2\pi^5 >1$. The latter occurs for $B\sim1.6$, i.e. $\rho \sim 0.93$. Hence, we see that the SNIC bifurcation occurs for values of $\rho$ very close from the threshold above which the solution is no longer defined $\rho = 1$. This justify that we keep $p\sim p_c$ at a threshold above which the motion stops, although the SNIC stops it slighlty before $p \sim 0.9 p_c$.

Direct numerical simulations presented in Fig. \ref{fig:S3}c-e show the SNIC bifurcation. Starting from DNS of Eq. \ref{eq:psiTot}, we compute the slow envelope and then the position of the kink using \eqref{eq:defXsoliton}. At low forcing, the velocity is constant $(f_1 \approx 0$). For higher forcing, one observes more variability of the velocity around each site, in qualitative agreement with prediction, although the quantitative agreement is not perfect. Last, when forcing exceeds a threshold, the kink remains stuck, indicating the bifurcation has been crossed.
\clearpage

\section{Experimental calibration}

In this section, we determine the parameters used in the discrete model $I\,\ddot{\theta}_i = - k_0 \theta_i - k_c \theta_i^3 - \gamma_0 \dot{\theta}_i - k_b \left( 2\theta_i - \theta_{i+1} - \theta_{i-1} \right) - \gamma_b \left( 2\dot{\theta}_i - \dot{\theta}_{i+1} - \dot{\theta}_{i-1} \right) + \epsilon \cos\left( 2\omega t + \phi_i \right)\theta_i$. The parameters to be calibrated are $I$ (moment of inertia), $k_0$ (onsite stiffness), $k_c$ (cubic stiffness), $k_b$ (coupling stiffness), $\gamma_0$ (onsite damping), and $\gamma_b$ (viscous coupling damping).

\subsection{Summary of calibrated parameters}

\begin{table}[h!]
\centering
\label{tab:calibrated_parameters}
\renewcommand{\arraystretch}{1.3}
\begin{tabular}{|c|c|c|c|}
\hline
\textbf{Parameter} & \textbf{Description} & \textbf{Value} & \textbf{Units} \\
\hline
$I$ & Rotor moment of inertia &
$5.60\times10^{-5}$ &
kg\,m$^2$ \\
\hline
$k_0$ & Linear onsite stiffness &
$0.01501 \pm 0.00220$ &
N\,m\,rad$^{-1}$ \\
\hline
$k_c$ & Cubic onsite stiffness &
$0.00903 \pm 0.00227$ &
N\,m\,rad$^{-3}$ \\
\hline
$k_b$ & Linear coupling stiffness &
$0.00429 \pm 0.00106$ &
N\,m\,rad$^{-1}$ \\
\hline
$\gamma_0$ & Onsite damping coefficient &
$(1.867 \pm 0.192)\times10^{-4}$ &
N\,m\,s\,rad$^{-1}$ \\
\hline
$\omega_0$ & Natural angular frequency &
$16.33 \pm 1.18$ &
rad\,s$^{-1}$ \\
\hline
\end{tabular}
\caption{Calibrated parameters used in the nonlinear discrete rotor model. Values are reported as mean $\pm$ standard deviation from the calibration ensemble.}
\end{table}
\begin{figure}[h!]
    \centering
    \includegraphics[width=0.8\linewidth]{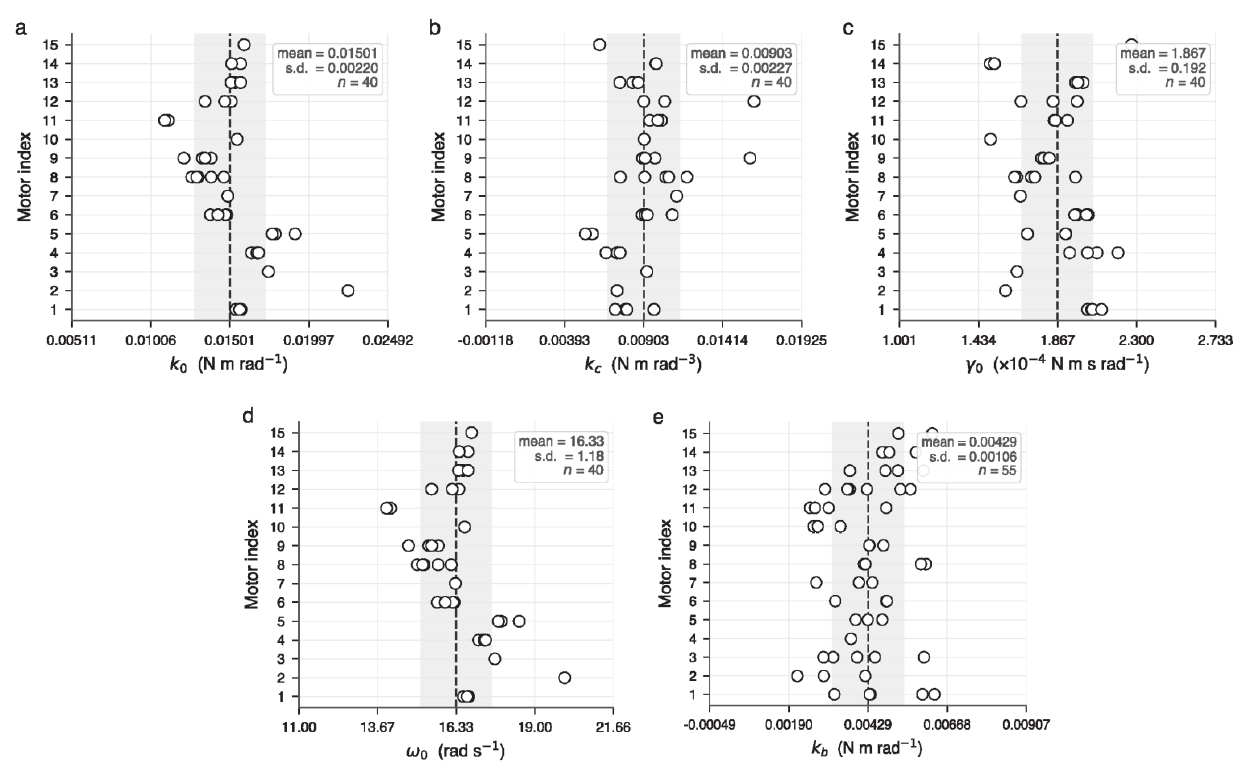}
    \caption{Calibration Data. Each line represents measurement on a different site. Standard deviation over all measurement is indicated by grey area.}
    \label{fig:placeholder}
\end{figure}
\subsection{Moment of inertia}

\begin{figure}
    \centering
    \includegraphics[width=0.7\linewidth]{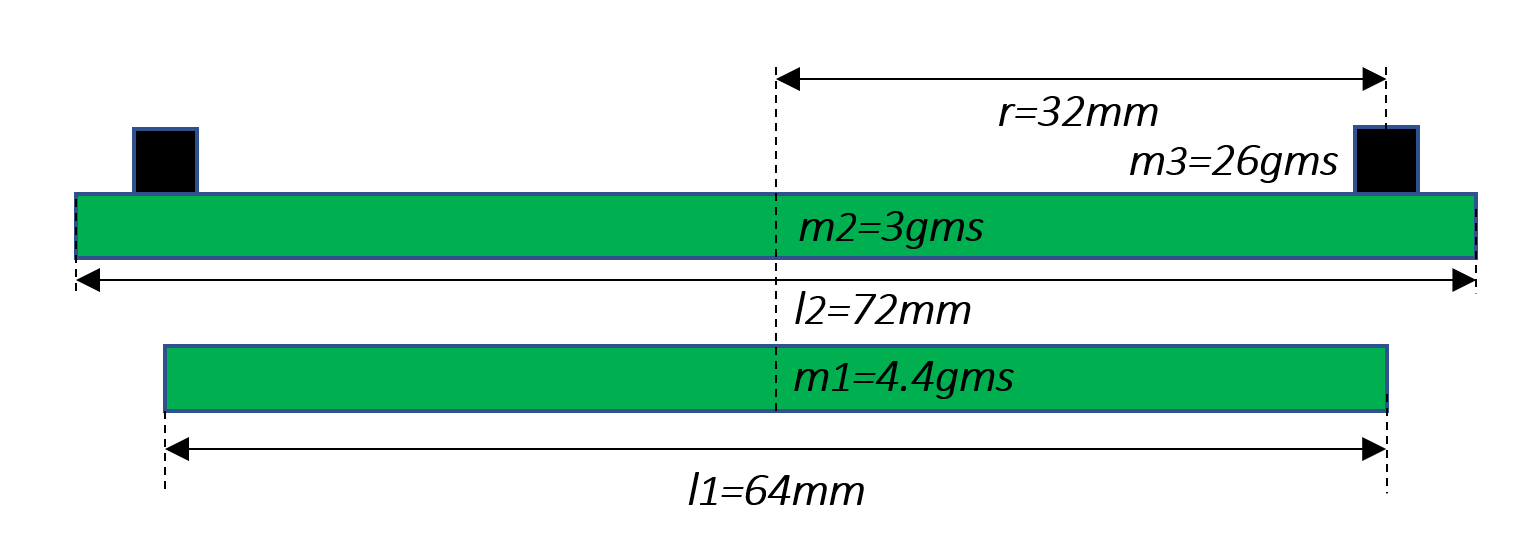}
    \caption{Geometry used for the calculation of the rotor moment of inertia.}
    \label{fig:inertia_calibration}
\end{figure}

The rotor moment of inertia is calculated by approximating the structure as two rods of lengths $l_1$ and $l_2$, together with two screw masses located at a distance $r$ from the center, as shown in Fig.~\ref{fig:inertia_calibration}.
The total moment of inertia is given by $I=\frac{m_1 l_1^2}{12}+\frac{m_2 l_2^2}{12}+2m_3 r^2$. Using SI units, $m_1 = 4.4\times10^{-3}\,\text{kg}$, $m_2 = 3.0\times10^{-3}\,\text{kg}$, $m_3 = 2.6\times10^{-2}\,\text{kg}$, $l_1 = 0.064\,\text{m}$, $l_2 = 0.072\,\text{m}$, and $r = 0.032\,\text{m}$. Substituting these values gives total moment of inertia is $\boxed{I = 5.60\times10^{-5}\,\text{kg}\cdot\text{m}^2}$.

\subsection{Stiffness and damping}

\begin{figure}
    \centering
    \includegraphics[width=1.0\linewidth]{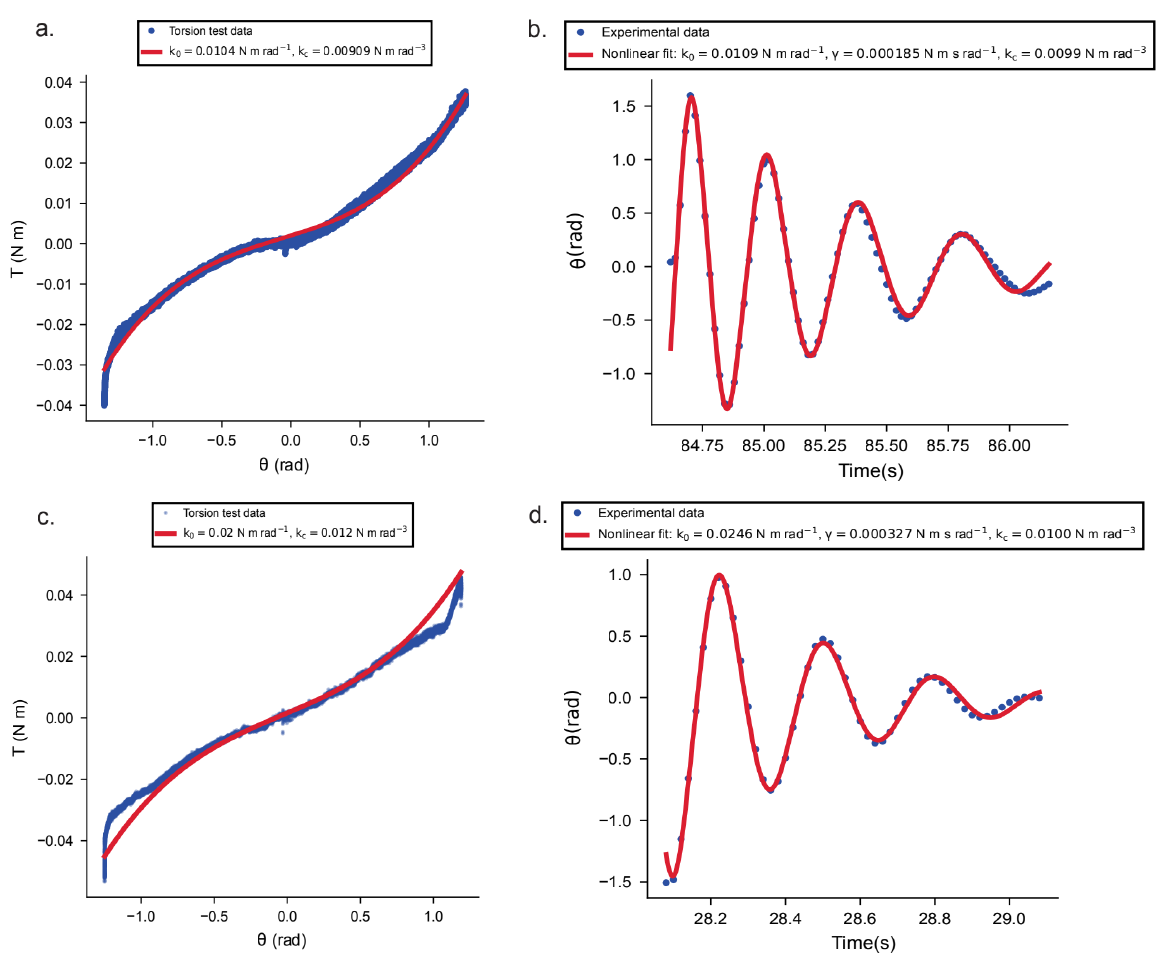}
    \caption{Calibration experiments used to determine onsite stiffness, coupling stiffness, cubic nonlinearity, and damping. Torque (Nm) on y axis and angular displacement(rad) on x axis. (a) Static torsion test with only onsite band. (b) Free oscillation fit for onsite response. (c) Static torsion test with onsite and coupling bands. (d) Free oscillation fit for the coupled system.}
    \label{fig:calibration_stiffness}
\end{figure}

The linear stiffness, cubic nonlinearity, and damping are determined using four calibration experiments shown in Fig.~\ref{fig:calibration_stiffness}(a--d).In Fig.~\ref{fig:calibration_stiffness}(a), a three-unit system is mounted on the Instron E3000 testing machine. The rotations of the two neighboring units are clamped and the coupling bands are removed, leaving only the onsite band. Angular displacement is imposed on the central unit and the resulting torque is measured. This gives the torque--angle response associated only with the onsite restoring force. A linear and a cubic fit are applied to the torsion curve, giving $k_0 = 0.0104 \ \text{N\,m\,rad}^{-1}$ and $k_c = 0.00909 \ \text{N\,m\,rad}^{-3}$. These values correspond to the linear onsite stiffness and the cubic onsite nonlinearity. In Fig.~\ref{fig:calibration_stiffness}(b), the same unit is removed from the Instron and subjected to a free perturbation experiment. The measured angular displacement $\theta(t)$ is fitted using $I\ddot{\theta}=-k_0\theta-\gamma_0\dot{\theta}-k_c\theta^3$. The fit gives $k_0 = 0.0109 \ \text{N\,m\,rad}^{-1}$, $\gamma_{01}= 1.85\times10^{-4} \ \text{N\,m\,s\,rad}^{-1}$, and $k_c = 0.0099 \ \text{N\,m\,rad}^{-3}$. These values are consistent with the static torsion measurement and confirm the onsite calibration. The damping coefficient $\gamma_{01}$ includes the contribution from shaft friction and the onsite elastic band.

In Fig.~\ref{fig:calibration_stiffness}(c), the same three-unit system is tested again, but now both coupling bands are connected together with the onsite band. The measured torque-
angle response therefore contains both onsite and coupling contributions. The linear fit gives the total stiffness $k_{\mathrm{total}} = 0.020 \ \text{N\,m\,rad}^{-1}$, while the cubic fit gives $k_c = 0.012 \ \text{N\,m\,rad}^{-3}$. Since the central unit is connected to two neighboring units, the total linear stiffness satisfies $k_{\mathrm{total}} = k_0 + 2k_b$, where $k_b$ is the stiffness contribution of one coupling band. Using the value of $k_0$ obtained from Fig.~\ref{fig:calibration_stiffness}(a), we obtain $k_b=\frac{k_{\mathrm{total}} - k_0}{2}=\frac{0.020 - 0.0104}{2}=0.0048 \ \text{N\,m\,rad}^{-1}$. This gives the linear coupling stiffness of a single coupling band.

In Fig.~\ref{fig:calibration_stiffness}(d), the coupled system is perturbed dynamically and the free oscillatory response is fitted using the same nonlinear equation. The fit gives $k_0 = 0.0246 \ \text{N\,m\,rad}^{-1}$, $\gamma = 3.27\times10^{-4} \ \text{N\,m\,s\,rad}^{-1}$, and $k_c = 0.0100 \ \text{N\,m\,rad}^{-3}$. This corresponds to the contribution friction from shaft, onsite stiffness, and coupling stiffness. The increase in both stiffness and damping compared to Fig.~\ref{fig:calibration_stiffness}(b) reflects the additional restoring force and dissipation introduced by the coupling bands. In contrast to the tests above, the oscillators in the experiments remain synchronized most of the time. Therefore, we assume that the contribution from the viscous damping term is negligible.
Together, these four measurements determine the parameters $k_0$, $k_b$, $k_c$, and $\gamma_0$ used in the nonlinear discrete model. 

For completeness, we remind once again here how to adimension the equation to perform numerical simulations. We start from
\begin{equation}
I\,\ddot{\theta}_i
=
- k_0 \theta_i
- k_c \theta_i^3
- \gamma_0 \dot{\theta}_i
- k_b \left( 2\theta_i - \theta_{i+1} - \theta_{i-1} \right)
+ \epsilon \cos\left( 2\omega t + \phi_i \right)\theta_i .
\end{equation}
We divide everything by I. We then define $\omega_0^2=k_0/I$ and set $\theta(t) = \hat \theta(\omega_0 \tau)$ such that $\dot \theta = \omega_0 \partial_t \hat{\theta}$. We obtain (removing hats)
\begin{equation}
\ddot{\theta}_i
=
-  \theta_i
- (k_c/I\omega_0^2)\theta_i^3
- (\gamma_0/I\omega_0) \dot{\theta}_i
- (k_b/I\omega_0^2) \left( 2\theta_i - \theta_{i+1} - \theta_{i-1} \right)
+ (\epsilon/I\omega_0^2) \cos\left( 2\omega t/\omega_0 + \phi_i \right)\theta_i .
\end{equation}
We then define $\phi=\sqrt{k_c/k_0} \theta$, and pick $\omega=\omega_0$. We obtain
\begin{equation}
\ddot{\phi}_i
=
-  \phi_i
- \phi_i^3
- \gamma_0/(I \omega_0) \dot{\phi}_i
- (k_b/k_0) \left( 2\phi_i - \phi_{i+1} - \phi_{i-1} \right)
+ (\epsilon/k_0) \cos\left( 2t + \phi_i \right)\phi_i .
\end{equation}
In numerical simulations, we use the form
\begin{equation}
\ddot{\phi}_i
=
- (1 + \hat \epsilon\cos(2t+\phi_i)) \phi_i
- \phi_i^3
- 2 \hat \Gamma \dot{\phi}_i
- \hat \tau \left( 2\phi_i - \phi_{i+1} - \phi_{i-1} \right)
\end{equation}
Hence the simulation parameters should be chosen as follow : 
\begin{equation}
    \hat \epsilon = \epsilon/k_0 \qquad \hat \Gamma = \gamma/(2 I\omega_0) = \gamma/2\sqrt{Ik_0} \qquad \tau = k_b/k_0
\end{equation}
and the result of the simulation $\hat \theta(t)$ relates to the actual physical variable $\theta$ by
\begin{equation}
    \theta(t) = \sqrt{\frac{k_c}{k_0}}\hat \theta(\omega_0 t)
\end{equation}
\begin{figure}[t]
    \centering
    \includegraphics[width=1\linewidth]{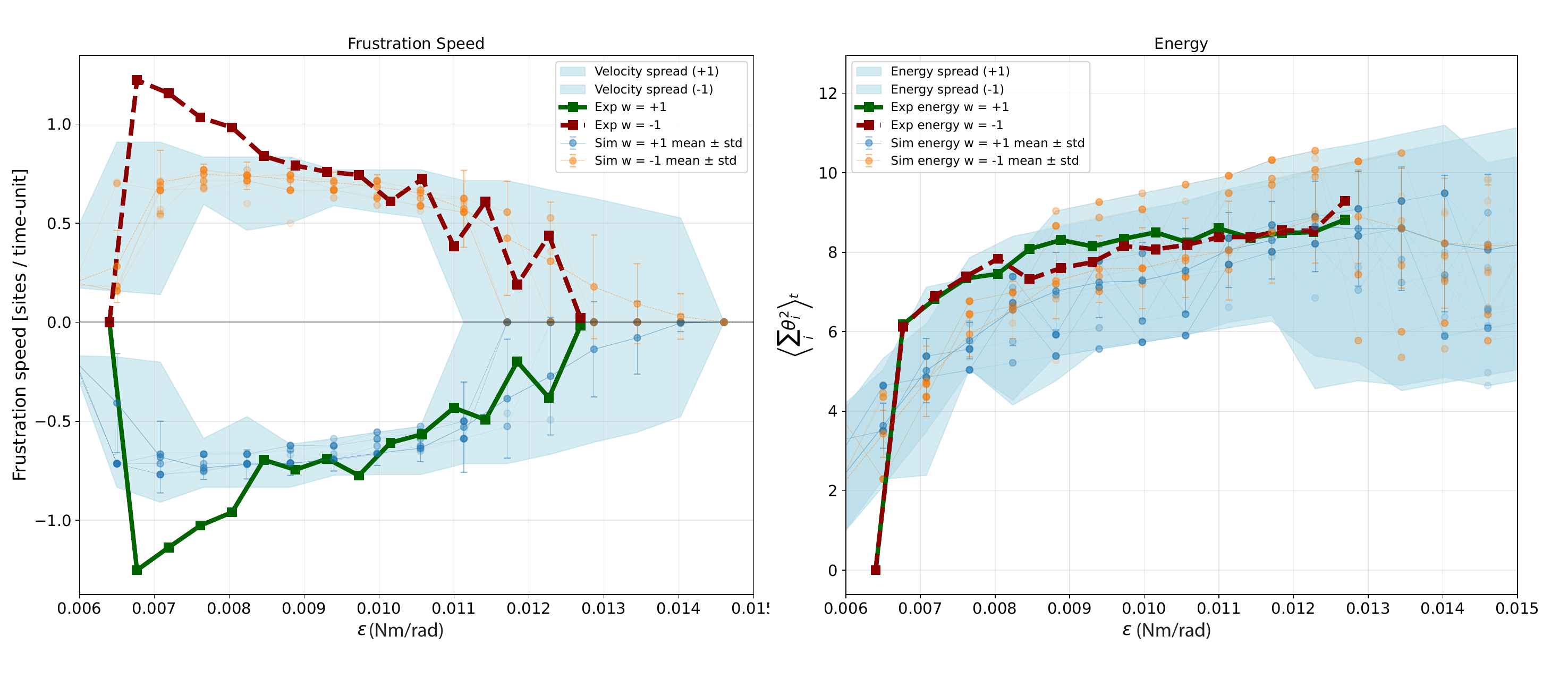}
    \caption{Comparison of frustration velocity and system energy between experiments and simulations. The shaded blue region indicates the range of simulation results obtained by considering a 10\% variation in the calibrated parameters. Experimental measurements fall within this range, demonstrating strong agreement between the simplified model and the observed dynamics.}
    \label{fig:exp_vs_theory}
\end{figure}

In Fig.~\ref{fig:exp_vs_theory}, we compare the frustration velocity and the system energy obtained from experiments and simulations. The simulations use the calibrated parameters extracted from the previous tests while accounting for a 10\% variation in these parameters, and for each value of driving amplitude $\varepsilon$ we ran 20 iterations ,along with it was done over 10 different initial conditions to capture experimental uncertainty. The system energy is calculated by averaging the square of the oscillator amplitude over both space and time. The light blue shaded region represents the range of frustration velocity (sites/sec) and system energy predicted by the simulations under this parameter variation. We observe that the experimental results lie well within these ranges, indicating good agreement between theory and experiments.Despite keeping the model intentionally simple by neglecting viscous damping and other higher-order nonlinearities, the model is still able to capture the observed phenomenon both qualitatively and quantitatively.
\section{Arnold's Tongue : Experiment ,theory and Numerical Simulation}

\subsection{Theoretical Prediction}

Again we come back to the equation $I \ddot{\theta} + \gamma \dot{\theta} + k_0 \theta + k_c \theta^3 = \epsilon \cos(\omega t)\,\theta$. To analyze the onset of parametric resonance, we first normalize the equation by dividing by $I$, giving $\ddot{\theta} + 2\beta \dot{\theta} + \omega_0^2 \theta + \alpha \theta^3 = h \cos(\omega t)\,\theta$, where $\omega_0^2 = \frac{k_0}{I}$, $2\beta = \frac{\gamma}{I}$, $\alpha = \frac{k_c}{I}$, and $h = \frac{\epsilon}{I}$. Neglecting nonlinearities ($k_c \to 0$), this reduces to a Mathieu-type equation, $\ddot{\theta} + 2\beta \dot{\theta} + \omega_0^2 \theta - h \cos(\omega t)\,\theta = 0$.

The principal parametric resonance occurs near $\omega \approx 2\omega_0$. Introducing $\nu = \frac{\omega}{2}$ and using the Floquet ansatz $\theta(t)=A e^{i\nu t}+B e^{-i\nu t}$, while retaining only the resonant terms, we obtain

\begin{equation}
\begin{pmatrix}
\omega_0^2 - \nu^2 + 2 i \beta \nu & -h/2 \\
-h/2 & \omega_0^2 - \nu^2 - 2 i \beta \nu
\end{pmatrix}
\begin{pmatrix}
A \\ B
\end{pmatrix}
= 0.
\end{equation}
\begin{figure}
    \centering
    \includegraphics[width=1.1\linewidth]{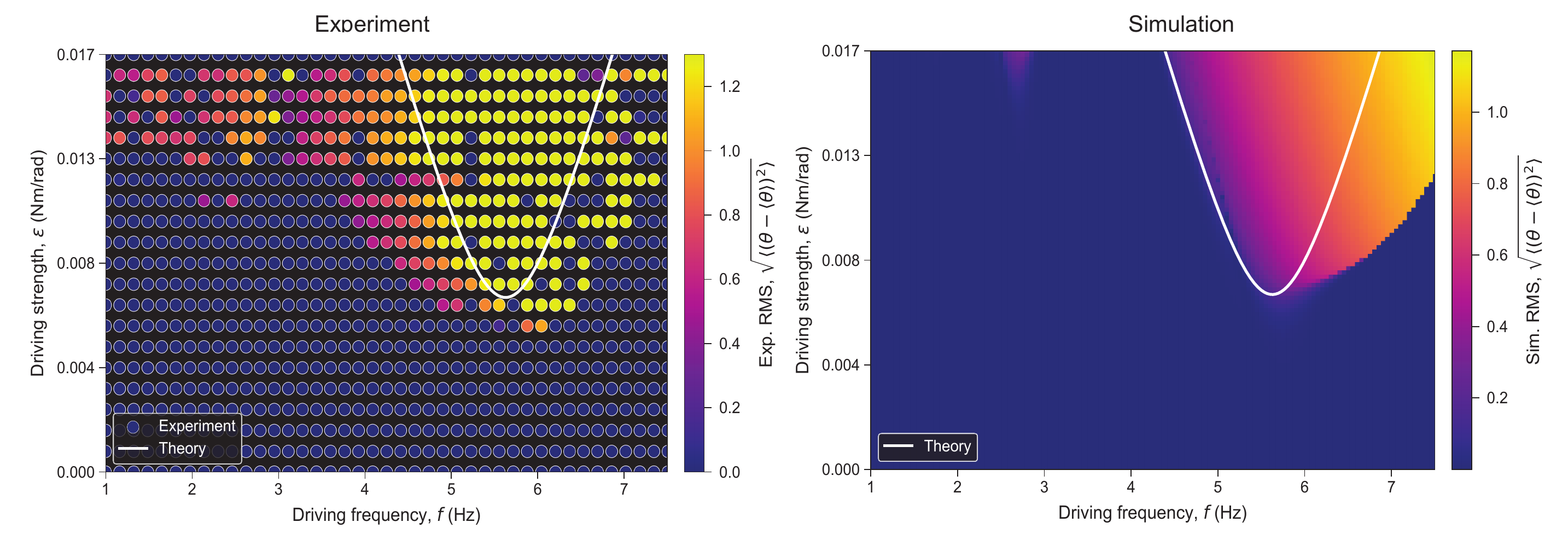}
    \caption{Arnold Tongue Experiment, Theory and Simulation with x axis as driving frequency (Hz) and y axis as Driving Amplitude(Nm/rad). RMS is plotted as colorplot}
    \label{fig:placeholder}
\end{figure}

A nontrivial solution exists only if the determinant vanishes, giving $\left(\frac{h}{2}\right)^2 = \left(\omega_0^2 - \nu^2\right)^2 + \left(2\beta \nu\right)^2$. Substituting back the original parameters gives the instability boundary (Arnold tongue),
\begin{equation}
\boxed{
\epsilon_c(\omega)
=
2I
\sqrt{
\left(
\omega_0^2
-
\frac{\omega^2}{4}
\right)^2
+
\left(
\frac{\gamma \omega}{2I}
\right)^2
}
}
\end{equation}
with $\omega_0 = \sqrt{\frac{k_0}{I}}$.The cubic stiffness term $k_c$ does not significantly modify the linear instability threshold, but it determines the finite-amplitude response inside the Arnold tongue. Using averaging or harmonic balance, the steady-state amplitude $r$ satisfies
$
\left(
\frac{\epsilon}{2I}
\right)^2
=
\left[
\omega_{0}^2
-
\frac{\omega^2}{4}
+
\frac{3k_c}{4I}r^2
\right]^2
+
\left[
\frac{\gamma\omega}{2I}
\right]^2.
$which gives,
$
r^2
=
\frac{4I}{3k_c}
\left[
\pm
\sqrt{
\left(
\frac{\epsilon}{2I}
\right)^2
-
\left(
\frac{\gamma\omega}{2I}
\right)^2
}
-
\left(
\omega_0^2
-
\frac{\omega^2}{4}
\right)
\right].
$This shows that $k_c > 0 \Rightarrow \text{hardening nonlinearity}$, which bends the resonance toward higher frequencies, while $k_c < 0 \Rightarrow \text{softening nonlinearity}$, which bends the resonance toward lower frequencies.This expression defines the Arnold tongue separating stable and unstable regions in the $(f, \epsilon)$ parameter space. We perform the experimental sweep of parametrically driven frequency and amplitude in a single unit with onsite stiffness (without coupling bands). For comparison with experiments, the driving amplitude $\epsilon_e$ (electronic) is related to $\epsilon$ via a proportionality factor $\epsilon = \alpha A$, where $\alpha$ accounts for the motor calibration. The calibrated value is $\alpha = 0.846 \,\text{Nm/rad}$.

\subsection{Numerical Simulation}
We numerically integrated the equation $
I \ddot{\theta} = 
- k_0 \theta 
- k_c \theta^3 
- \gamma \dot{\theta} 
+ \epsilon \cos(\omega t)\,\theta,
$ over a grid of driving amplitudes and frequencies. After discarding an initial transient, the steady-state response is quantified using the root-mean-square (RMS) amplitude :
$
\theta_{\mathrm{RMS}}
= \sqrt{
\left\langle
\left(\theta - \langle \theta \rangle \right)^2
\right\rangle
}.
$ This measure captures the oscillation strength while removing any static offset. The resulting RMS values are used to construct numerical Arnold tongues for direct comparison with theory and experiment. We observe good agreement between experiments and simulations, while the linear theory accurately predicts the key charecterstic and primary instability tip near 2f, corresponding to the lower threshold of instability set by dissipation.
\clearpage
\section{Additional experimental results with respect to winding}
\begin{figure}
    \centering
    \includegraphics[width=0.8\linewidth]{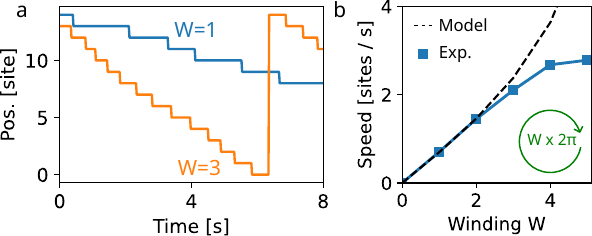}
    \caption{(a) Experimental trajectories for different winding values $W=1,3$, forcing being fixed to $\epsilon=0.8$. (b) Velocity of frustration as function of winding $W$ and theory with no adjustable parameter.}
    \label{fig:placeholder}
\end{figure}

As discussed in the main text, the velocity also depends on the winding number according to Eq. (\ref{eq:velocityLowForcing}). The dislocation speed increases due to faster energy transfer, related to larger phase difference between neighbors. We propose to test this prediction by driving the chain with constant pump amplitude pump and various values of $W \in \mathbb N$. Typical trajectories of the dislocation are shown in Fig. 7a, confirming faster motion at higher winding. For $W=1$, the model predicts a velocity of $1/\tau=k_b/\sqrt{k_0 I}=1.5$ site/s using calibrated values from above, which is of the correct order of magnitude but about two times larger than the one actually observed (Fig. 7b). This suggests that $f(p)$ is not exactly zero in our experiments, but slightly larger and slows down the dislocation. We therefore use a modified expression for the velocity with respect to winding
\begin{equation}
    v_0 =  - \frac{Ck}{\omega_0} \sin\left(\frac{\pi W}{N} \right)
\end{equation}
where $C=0.5$ calibrated from $W=1$ measurement accounts for $f(p) >0$. The dependency of velocity with respect to winding is well captured by the model (dashed line in Fig. 7b). The direction and velocity of the propagation can thus be controlled by tuning the sign and magnitude of phase difference between neighboring sites.

\bibliographystyle{unsrt}
\bibliography{biblio}